\documentclass[12pt,a4paper,BCOR10mm]{article}
\usepackage[french]{babel}
\usepackage{graphicx}
\usepackage{times}
\usepackage{amsmath}
\usepackage{float}
\usepackage{blindtext}
\usepackage{amsfonts}
\usepackage[latin1]{inputenc}  
\title{Nilpotent Morse algebra and time evolution of certain associated coherent states}
\author{A. Belfakir$^{a,}$ \footnote{abdobelfakir01@gmail.com} ,$\hspace{0.3 cm}$
Y. Hassouni$^{a,b,}$ \footnote{yassine.hassouni@mnf.uni-tuebingen.de}
\\  \small $^{a}$ Equipe des Sciences de la matière et du Rayonnement, ESMAR \\\small Faculté des sciences, Université Mohammad V-Agdal, Avenue Ibn Batouta, B.P.1014,\\ \footnotesize Agdal , Rabat-Morocco \\  \small $ ^{b} $ Institut f\"ur theoretische physik, Universit\"at t\"ubingen, 72076 T\"ubingen, Germany}

\date{}

\begin{document}

\maketitle
\section*{Abstract}
We provide the time evolutions of the linear and nonlinear coherent states for several systems characterized by different energy spectra, and we identify the regions in the parameter
space where these systems behave closer to the classical systems. The Morse system is algebraically found within the frame of  generalized Heisenberg algebra(GHA). We demonstrate that this system is described by a nilpotency condition.  Then, we propose a construction of coherent states for the Morse oscillator.\\
Keywords: Generalized Heisenberg algebra (GHA), Coherent states(CS), Morse potential.
 \section{Introduction}
    
    Algebraic methods have played an important role in describing the quantum physical systems from the early days of quantum mechanics (harmonic oscillator algebra described by the creation and annihilation operators). Later the deformed Heisenberg algebras[1] were constructed and applied to many areas of physics, such as nuclear physics [2], and condensed matter [3].\\
    In the last decade, the generalized Heisenberg algebra (GHA)[4] has been constructed and applied to several systems as a system in an infinite square-well potential[5-6], and the Hydrogen atom[7]. This algebra is described through the generators $J_{0}$ , $A$ and $A^{\dagger}$, where  $J_{0}$ is a hermitian operator, and $A$, $A^{\dagger}$ are called the ladder operators of the physical system under consideration. These generators are related to
each other by a general function called the characteristic function of the algebra which relies on two  successive energy levels $\varepsilon_{n}$ and $\varepsilon_{n+1}$, such that $\varepsilon_{n+1}=f(\varepsilon_{n})$. In other words, any physical system whose  energy eigenvalues obey $\varepsilon_{n+1}=f(\varepsilon_{n})$, can be described by the generalized Heisenberg algebra(GHA) with its characteristic function $f(x)$. It is important to mention that the generalized Heisenberg algebra contains the harmonic oscillator
and q-deformed algebras as particular cases. \\
 On the other hand, there is a wide consensus that coherent states are the most classical states, These  states are useful for investigating various problems in different areas of physics[8]. These are states that have the property of minimizing the Heisenberg uncertainty relation. They were first introduced by Shr$\ddot{o}$dinger in the context of the harmonic oscillator, when he was searching for the connection between quantum physics and classical physics [9]. Then, Glauber applied some of the properties of coherent states in description of
statistical light beams[10]. Next, the $SU(2)$ and $SU(1;1)$ coherent states were introduced by Perelomov. These have been associated with several quantum systems in many areas of physics such as quantum optics, statistical mechanics and nuclear physics[11-12]. We would like to note that there are different approaches to constructing these coherent states(Klauder's approach and Perelomov's approach)[13-14].\\
Recently, the canonically conjugate operators for the generalized Heisenberg algebra(GHA) were introduced, which are associated with the generalized harmonic oscillator creation and annihilation operators. These generalized operators obey the same algebra as the harmonic oscillator ladder operators. The corresponding coherent states were constructed and are called the linear coherent states[15].\\
In sections $2$ and $3$ we give a brief overeview of GHA and the
generalized creation and annihilation operators. In section 4 we recall the definitions of the GHA coherent state and the linear coherent state. Subsequently, in section $5$ we calculate the time evolution of the uncertainty relation of the GHA coherent states and the linear coherent states for particular systems. 
    In section $6$ we provide the GHA of the one dimensional Morse oscillator, whose
Shr$\ddot{o}$dinger equation solutions are well known, by finding the corresponding characteristic function. Then, we show that GHA, in this case, is a nilpotent algebra, so the representation is finite. Finally, we present the expression of the coherent state corresponding to the Morse oscillator, and we calculate the time evolution of the uncertainty relation on this coherent state.   \\
   
    \section{Review of Generalized Heisenberg Algebra}
    First, let us begin by a brief review of GHA detailed in [4]. Indeed, GHA is described by the generators $J_{0}$, $A$, and $A^{\dagger}$ where $A=(A^{\dagger})^{\dagger}$ and $J_{0}$($J_{0}=J_{0}^{\dagger}$) is a hermitian operator. In [5] the operator $J_{0}$ was taken as the Hamiltonian of the physical systems under consideration. These generators $J_{0}$, $A$, and $A^{\dagger}$ satisfy the following relations 
    \begin{equation}
    J_{0}A^{\dagger}=A^{\dagger}f(J_{0}),
    \end{equation}
    \begin{equation}
   AJ_{0}=f(J_{0})A,
    \end{equation}
    \begin{equation}
  [A,A^{\dagger}]=f(J_{0})-J_{0},
    \end{equation}
    where $f$ is an analytical function of $J_{0}$ called the characteristic function of the algebra. This function depends on the physical system under consideration. A quantum system can be described by GHA by choosing the appropriate characteristic function, e.g the harmonic oscillator is recognized by taking $f(x)=x+1 $ as the characteristic function of GHA [4], and GHA ,in this case, becomes the ordinary Heisenberg algebra
 \begin{equation}
   HA^{\dagger}=A^{\dagger}H+A^{\dagger},
    \end{equation}
    \begin{equation}
   AH=HA+A,
    \end{equation}
    \begin{equation}
  [A,A^{\dagger}]=I,
    \end{equation} where $H$ is the Hamiltonian of the Harmonic oscillator ($J_{0}=H$), and $I$ is the identity operator.\\Consequently, GHA may be seen as a generalization of Heisenberg algebra, for this reason, it has been called GHA.\\
   Another example that may be seen as a specific case of GHA is the deformed Heisenberg algebra, such that the deformed oscillator is described by the charateristic function  $f(x)=q x+1$, where $q$ is the deformed parameter[1]. It is well known that these deformed algebras are associated with fractional statistics. \\Let us mention, another example totally described by GHA,  the free particle in a square-well potential $V(x)=0$ for $0<x<L$ and $V(x)=\infty$ elsewhere [5], its charateristic function is $f(x)=(\sqrt{x}+\sqrt{b})^{2}$ , where $b$ is a constant that characterizes the system. The GHA, in this case, becomes 
    \begin{equation}
   [H,A^{\dagger}]=2\sqrt{b}A^{\dagger}\sqrt{H}+bA^{\dagger},
   \end{equation}
   \begin{equation}
   [H,A]=-2\sqrt{b}\sqrt{H}A-bA,
   \end{equation}
    \begin{equation}
   [A,A^{\dagger}]=2\sqrt{b}\sqrt{H}A+b.
   \end{equation}
  In this case, the generator $J_{0}$ of GHA is the Hamiltonian $H$ of the system.  \\
  Now we give a concise review on the Fock representation of GHA. The basis of the Fock space associated with this algebra is the set  $\lbrace\left\vert n \right\rangle , n=0, 1,....\rbrace$ such that 
  \begin{equation}
  J_{0} \left\vert n \right\rangle=\varepsilon_{n}\left\vert n \right\rangle.
  \end{equation}
 The main property of the characteristic function $f$ is that, after calculations, it satisfies  \begin{equation}
  \varepsilon_{n}=f^{(n)}(\varepsilon_{0}). 
\end{equation}  It is seen as an $n$-th iterate of  $\varepsilon_{0}$, where $ \varepsilon_{0} $ is the eigenvalue of the generator $J_{0}$ corresponding to the vacuum state $\left\vert 0 \right\rangle$. In other words, this function connects two successive energy levels $\varepsilon_{n} $ and $\varepsilon_{n+1} $
\begin{equation}
 \varepsilon_{n+1}=f(\varepsilon_{n}).
\end{equation}
The operators $A$ and $A^{\dagger}$ act on a vector $\left\vert n \right\rangle$ as follows 
\begin{equation}
A^{\dagger}\left\vert n \right\rangle=N_{n}\left\vert n+1 \right\rangle,
\end{equation}
\begin{equation}
A\left\vert n \right\rangle=N_{n-1}\left\vert n-1 \right\rangle,
\end{equation}
where \begin{equation}
 N^{2}_{n}=\varepsilon_{n+1}-\varepsilon_{0}=f(\varepsilon_{n})-\varepsilon_{0}.
\end{equation}
The Casimir operator of GHA is given by\begin{equation}
\Gamma=AA^{\dagger}-f(J_{0}),
\end{equation}
this operator commute with all GHA operators $J_{0}$, $A$  and  $A^{\dagger}$.
\section{Generalized harmonic oscillator creation and annihilation operators}
It has been introduced in [15], that the GHA ladder operators can be associated with the operators $D$, $D^{\dagger}$ having the same algebraic structure as the harmonic oscillator ladder operators $a$ and $a^{\dagger}$. These are called the generalized harmonic oscillator creation and annihilation operators respectively, and it has been shown that 
\begin{equation}
 D=\sqrt{N+1}\frac{1}{\sqrt{f(H)-\varepsilon_{0}}}A ,
\end{equation}
where $A$ is the GHA annihilation operator, $f$ is the characteristic function of the system under consideration, and $N$ is the particle number operator ($N\left\vert n \right\rangle=n\left\vert n \right\rangle$).   We note that $D^{\dagger}$ is the hermitian conjugate of $D$. It is evident that the operators $D$ and $D^{\dagger}$ act on a Fock space vector $\left\vert n \right\rangle$ as 
\begin{equation}
 D\left\vert n \right\rangle=\sqrt{n}\left\vert n-1 \right\rangle,
 \end{equation} 

 \begin{equation}
 D^{\dagger}\left\vert n \right\rangle=\sqrt{n+1}\left\vert n+1 \right\rangle.
 \end{equation}
It is then easily be shown that the operators $D$ and $D^{\dagger}$, with the operator $N$ obey 
  \begin{equation}
 [N,D]=-D,
 \end{equation}
  \begin{equation}
 [N,D^{\dagger}]=D^{\dagger},
 \end{equation}
 \begin{equation}
 [D,D^{\dagger}]=I.
 \end{equation}
Let us recall that the operators $D$ and $D^{\dagger}$ correspond to the canonically conjugate  position-like and momentum-like operators($\xi$,$\rho$) such that
 \begin{equation}
D=\frac{1}{\sqrt{2}}(\frac{\xi}{L}+\frac{iL}{\hbar}\rho),
\end{equation}
\begin{equation}
D^{\dagger}=\frac{1}{\sqrt{2}}(\frac{\xi}{L}-\frac{iL}{\hbar}\rho),
\end{equation}
where $L$ is a constant which has the dimension of the length. The position-like and momentum-like operators are given in terms of $D$ and $D^{\dagger}$ as 
\begin{equation}
\xi=\frac{L}{\sqrt{2}}(D+D^{\dagger}),
\end{equation}
\begin{equation}
\rho=i\frac{\hbar}{\sqrt{2}L}(D^{\dagger}-D),
\end{equation}
we note that  ($\xi$,$\rho$) have the interesting property $[\xi,\rho]=I$.
\section{Linear and Nonlinear  coherent states}
Let us recall that the GHA coherent state (the nonlinear coherent state) was introduced as an eigenstate of the  GHA annihilation operator  $A\left\vert z \right\rangle=z\left\vert z \right\rangle$, where $z$ is a complex number. It has been shown in [5] that the GHA coherent state can be written as
\begin{equation}
\left\vert z \right\rangle=N(|z|)\sum_{n=0}^{\infty}\frac{z^{n}}{N_{n-1}!}\left\vert n \right\rangle,
\end{equation}
where $N(|z|)$ is the normalization function . It was shown that this state satisfies the minimum set of conditions required to obtain Klauder's coherent state:\\
 i)normalizability, \\
 ii)the continuity in the label
  \begin{equation}
| z\longrightarrow z'|, ||\left\vert z \right\rangle \longrightarrow \left\vert z' \right\rangle||,
 \end{equation}
iii) completeness \begin{equation}
\int d^2z w(z)\left\vert z \right\rangle\left\langle z \right\vert=I.
\end{equation}
The linear coherent state has been defined as the eigenstate of the annihilation operator 
\begin{equation}
D \left\vert z \right\rangle_{L}=z \left\vert z \right\rangle_{L},
\end{equation}
and it can be written as
\begin{equation}
\left\vert z \right\rangle_{L}=e^{-\frac{|z|^2}{2}}\sum_{n=0}^{\infty}\frac{z^{n}}{\sqrt{n!}}\left\vert n \right\rangle.
\end{equation}
For the sake of simplification of calculations[15], from now on, we will take that $D\left\vert n \right\rangle=\sqrt{n-1}\left\vert n -1\right\rangle$ , for $n\geq 1$. Following this approach, the linear coherent state can be written as \begin{equation}
\left\vert z \right\rangle_{L}=e^{-\frac{|z|^2}{2}}\sum_{n=1}^{\infty}\frac{z^{n-1}}{\sqrt{(n-1)!}}\left\vert n \right\rangle,
\end{equation}
in the following we will take $z=r e^{i\varphi}$.\\
The time evolution of the states is obtained by the application of the unitary operator
\begin{equation}
U(t)=e^{-i\frac{Ht}{\hbar}},
\end{equation}
where $H$ is the Hamiltonian of the system under consideration.
 \section{Time evolution of the uncertainty relation $\Delta\xi\Delta\rho$ for GHA and linear coherent states of particular systems}
 Let us now calculate the time  evolution of the uncertainty relation $\Delta\xi\Delta\rho$ 
 \begin{equation}
\Delta\xi(t)\Delta\rho(t)=\sqrt{(<(\xi(t))^{2}>-(<\xi(t)>)^{2})(<(\rho(t))^{2}>-(<\rho(t)>)^{2})},
\end{equation}
on the linear and nonlinear coherent states of a simple class of spectra and on the linear and nonlinear coherent states of the Hydrogen atom.
 \subsection{Spectrum type 1}
 Let us consider a system whose energy spectrum is given by the expression
 \begin{equation}
 \varepsilon_{n}=b\frac{n}{n+1},
 \end{equation}
where $n\geq0$ and $b$ is a constant which has the dimension of energy, the corresponding GHA coherent state is given in [5] as
\begin{equation}
\left\vert r,\varphi \right\rangle=N(r)\sum_{n=1}^{\infty}\sqrt{n}r^{(n-1)}e^{i(n-1)\varphi}\left\vert n \right\rangle,
\end{equation}
where $N(r)=(1-r^2)$  and $0\leq r\leq 1 $.
\subsubsection*{Time evolutions of the mean values of $\xi$,$\rho$,$\xi^{2}$ and $\rho^{2}$ for the GHA coherent state}
The time evolutions of
the mean values of the operators $\xi$ and $\rho$ on the GHA coherent state (Eq. (36)) are
\begin{align}
<\xi(t)>&=\left\langle  r,\varphi \right\vert U^{\dagger}(t)\xi U(t) \left\vert r,\varphi \right\rangle\\
&=\frac{L}{\sqrt{2}}\left\langle  r,\varphi \right\vert U^{\dagger}(t)(D+D^{\dagger}) U(t) \left\vert r,\varphi \right\rangle\\&={L\sqrt{2}}(N(r))^{2}\lbrace \sum_{n=1}^{\infty}r^{2n-1}n\sqrt{(n+1)}\cos((\frac{n}{n+1}-\frac{n+1}{n+2})b\frac{t}{\hbar}+\varphi)\rbrace,
\end{align}
and
\begin{align}
<\rho(t)>&=\left\langle  r,\varphi \right\vert U^{\dagger}(t)\rho U(t) \left\vert r,\varphi \right\rangle\\
&=i\frac{\hbar }{L\sqrt{2}}\left\langle  r,\varphi \right\vert U^{\dagger}(t)(D^{\dagger}-D) U(t) \left\vert r,\varphi \right\rangle\\&=\frac{\sqrt{2}\hbar}{L}(N(r))^{2}\lbrace \sum_{n=1}^{\infty}r^{2n-1}n\sqrt{(n+1)}\sin((\frac{n}{n+1}-\frac{n+1}{n+2})b\frac{t}{\hbar}+\varphi)\rbrace,
\end{align}

Now, we calculate the time evolutions of the mean values of the operators $\xi^{2}$ and $\rho^{2}$ on the GHA coherent state (Eq. (36))
\begin{align}
<\xi(t)^{2}>&=\left\langle  r,\varphi \right\vert U^{\dagger}(t)\xi^{2} U(t) \left\vert r,\varphi \right\rangle\\
&=\frac{L^{2}}{2}\left\langle  r,\varphi \right\vert U^{\dagger}(t)(D^{2}+(D^{\dagger})^{2}+DD^{\dagger}+D^{\dagger}D) U(t) \left\vert r,\varphi \right\rangle 
\end{align} 
\begin{align}
\begin{split}
\hspace{4cm}=&L^{2}(N(r))^{2}\lbrace \sum_{n=1}^{\infty}r^{2n}n\sqrt{(n+1)(n+2)}\cos((\frac{n}{n+1}-\frac{n+2}{n+3})b\frac{t}{\hbar}+2\varphi) \\&\quad+\sum_{n=1}^{\infty}r^{2(n-1)}n(n-1)\rbrace +\frac{L^{2}}{2},
\end{split}
\end{align}

\begin{align}
<\rho(t)^{2}>&=\left\langle  r,\varphi \right\vert U^{\dagger}(t)\rho^{2} U(t) \left\vert r,\varphi \right\rangle\\
&=-\frac{\hbar^{2}}{L^{2}}\left\langle  r,\varphi \right\vert U^{\dagger}(t)(D^{2}+(D^{\dagger})^{2}-DD^{\dagger}-D^{\dagger}D) U(t) \left\vert r,\varphi \right\rangle
\end{align}
\begin{align}
\begin{split}
\hspace{3.5cm}=& -\frac{\hbar^{2}}{L^{2}}(N(r))^{2}\lbrace \sum_{n=1}^{\infty}r^{2n}n\sqrt{(n+1)(n+2)}\cos((\frac{n}{n+1}-\frac{n+2}{n+3})b\frac{t}{\hbar}+2\varphi)\\&\quad-\sum_{n=1}^{\infty}r^{2(n-1)}n(n-1)\rbrace+\frac{\hbar^{2}}{2L^{2}}.
\end{split}
\end{align}
Then, we can calculate easily the time evolution of the uncertainty relation Eq. $(34)$ for the GHA coherent state.\\ The figure 1 shows the time evolution of the uncertainty relation for the GHA coherent state of the particular system $1$ for two different space parameters $r=0.1$ and $r=0.5$ when $\frac{b}{\hbar}=1$ and $\varphi=0$. Looking at figure 1 we can see that the uncertainty oscillates between $0.5\hbar$
and $0.5020\hbar$ for $r=0.1$, however it oscillates between $0.5\hbar$ and $1.1\hbar$ when $r=0.5$.  As a consequence, the GHA coherent state is more stable for small parameters space $r$.
\begin{center}    
 \begin{figure}[H]
  \includegraphics[scale=0.5]{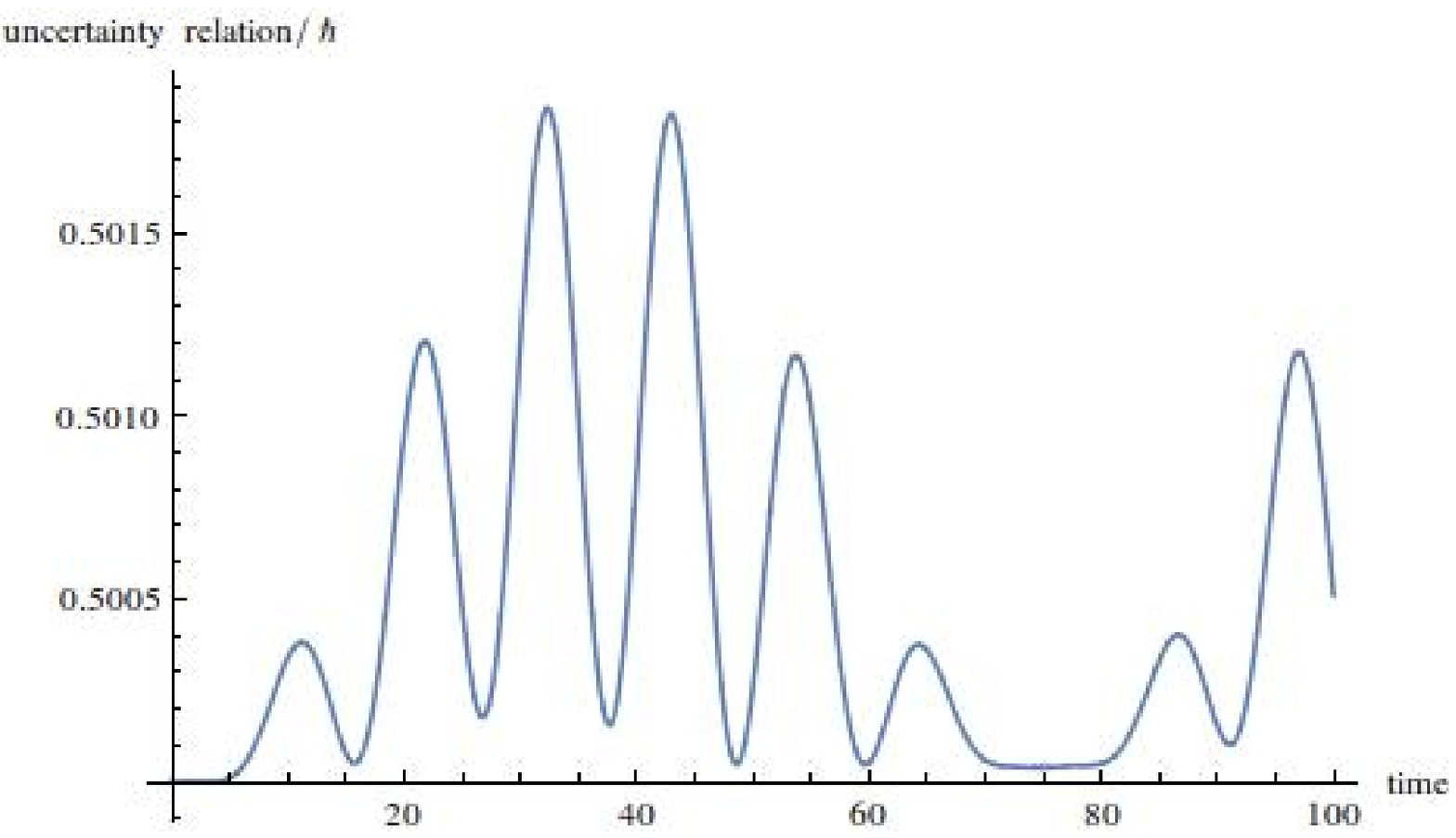}
 \includegraphics[scale=0.5]{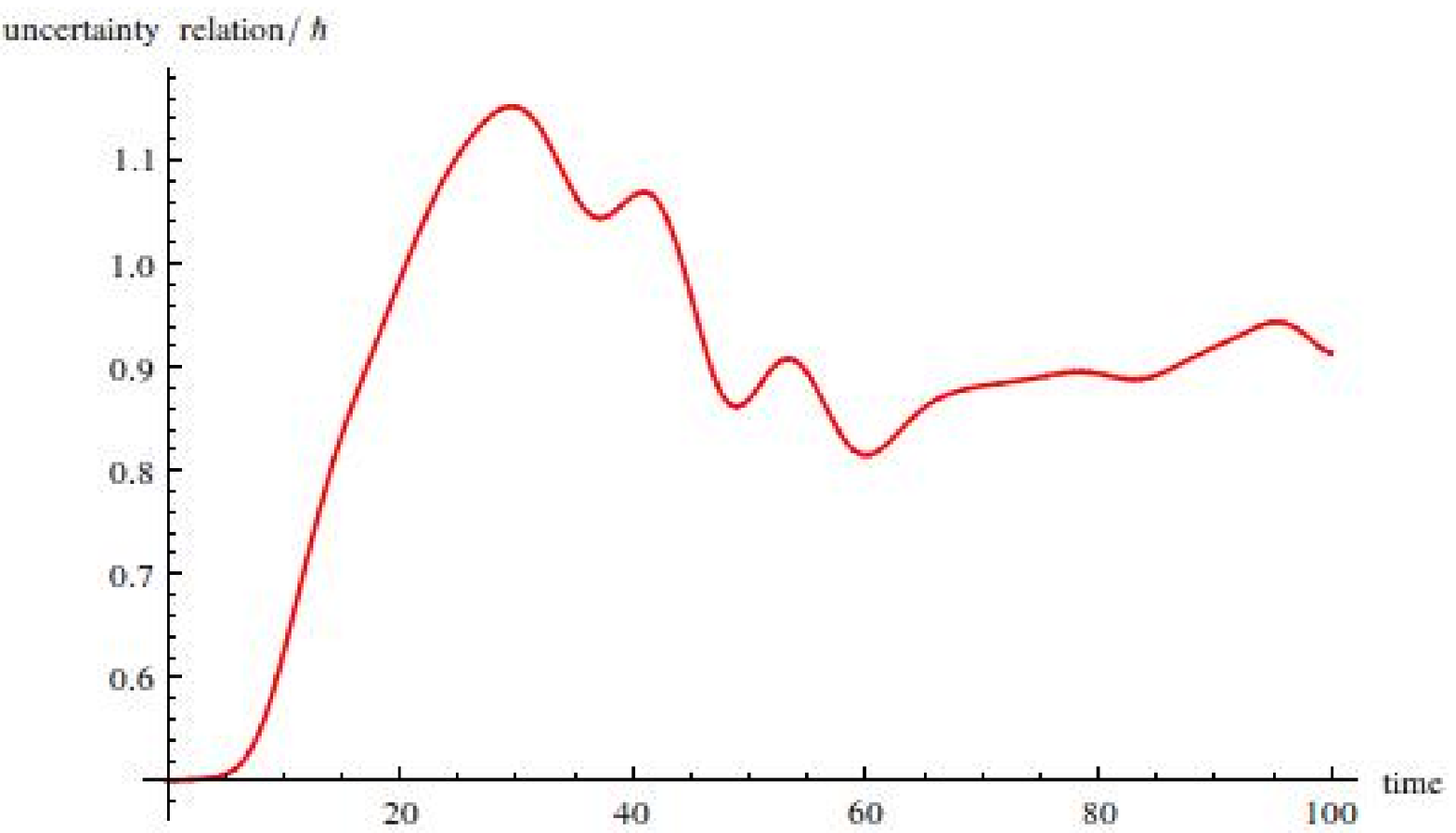}
  \caption{ The time evolution of the uncertainty relation  $\Delta\xi\Delta\rho/\hbar $ for $r=0.1$(blue curve) and $r=0.5$ (red curve) for GHA coherent states of system 1. in the two graphs we take $\varphi=0$ and $\frac{b}{\hbar}=1$.  }
 \end{figure} \end{center}
\subsubsection*{Time evolutions of the mean values of $\xi$,$\rho$,$\xi^{2}$ and $\rho^{2}$ for the linear coherent state}

The respective time evolutions of the mean values of the operators $\xi$, $\rho$, $\xi^{2}$ and $\rho^{2}$ on the linear coherent state( Eq. $(32)$) associated with the system whose energy spectrum is given by Eq. ($35$) are
 \begin{equation}
<\xi(t)>={L\sqrt{2}}e^{-r^{2}}\lbrace \sum_{n=1}^{\infty}\frac{r^{2n-1}}{(n-1)!}\cos((\frac{n}{n+1}-\frac{n+1}{n+2})b\frac{t}{\hbar}+\varphi)\rbrace,
\end{equation}
\begin{equation}
<\rho(t)>=\frac{\sqrt{2}\hbar}{L}e^{-r^{2}}\lbrace \sum_{n=1}^{\infty}\frac{r^{2n-1}}{(n-1)!}\sin((\frac{n}{n+1}-\frac{n+1}{n+2})b\frac{t}{\hbar}+\varphi)\rbrace,
\end{equation}
\begin{equation}
<\xi(t)^{2}>=L^{2}\lbrace e^{-r^{2}} \sum_{n=1}^{\infty}\frac{r^{2n}}{(n-1)!}\cos((\frac{n}{n+1}-\frac{n+2}{n+3})b\frac{t}{\hbar}+2\varphi)+r^{2}\rbrace+\frac{L^{2}}{2},
\end{equation}
\begin{equation}
<\rho(t)^{2}>= -\frac{\hbar^{2}}{L^{2}}\lbrace e^{-r^{2}} \sum_{n=1}^{\infty}\frac{r^{2n}}{(n-1)!}\cos((\frac{n}{n+1}-\frac{n+2}{n+3})b\frac{t}{\hbar}+2\varphi)-r^{2}\rbrace+\frac{\hbar^{2}}{2L^{2}}.
\end{equation}
Thus, by using Eqs. $(49)-(52)$ we can calculate the time evolution of the uncertainty relation Eq. $(34)$ for the linear coherent state.
\begin{center}
\begin{figure}[H]
  \includegraphics[scale=0.5]{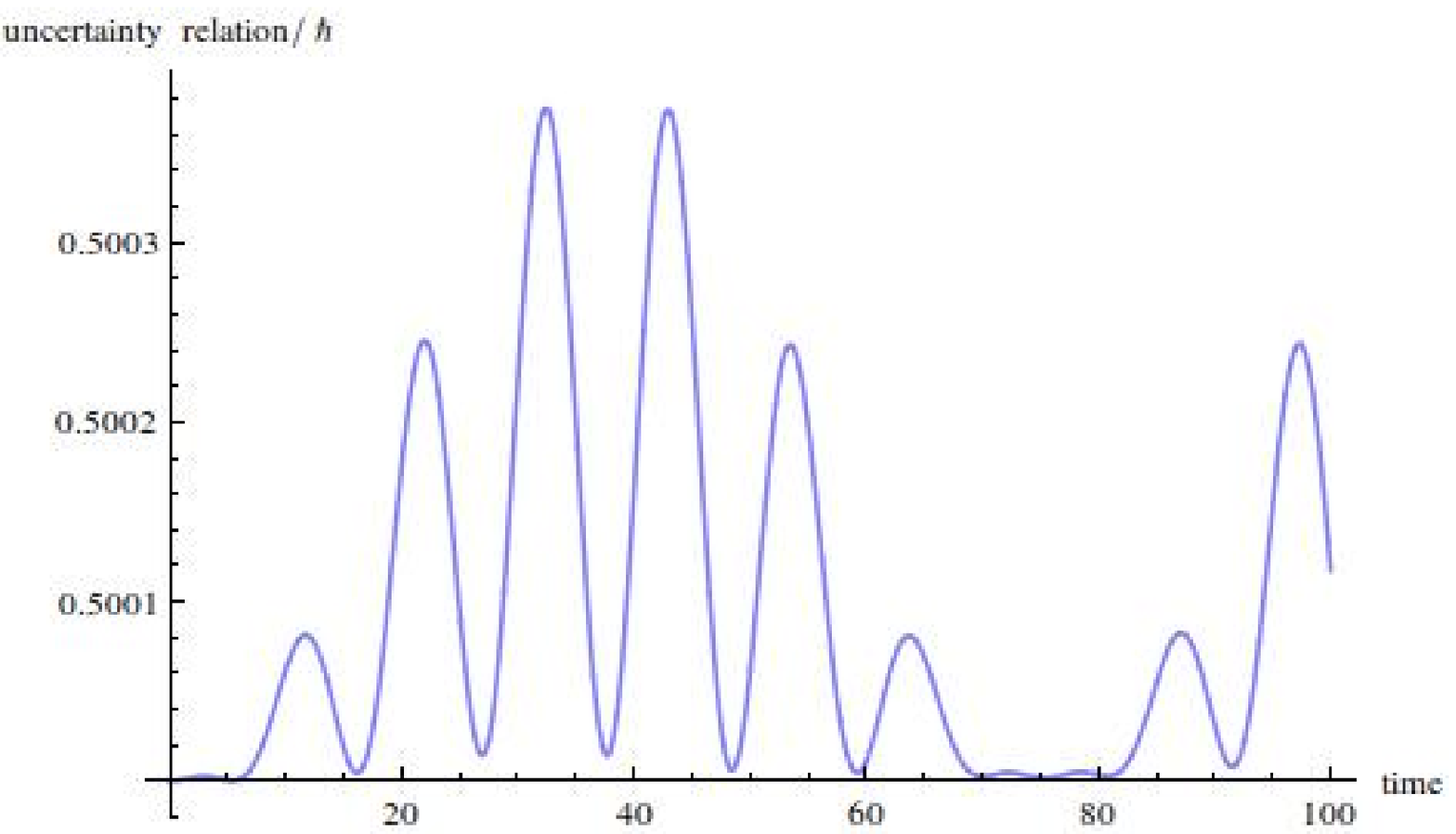}  
 \includegraphics[scale=0.5]{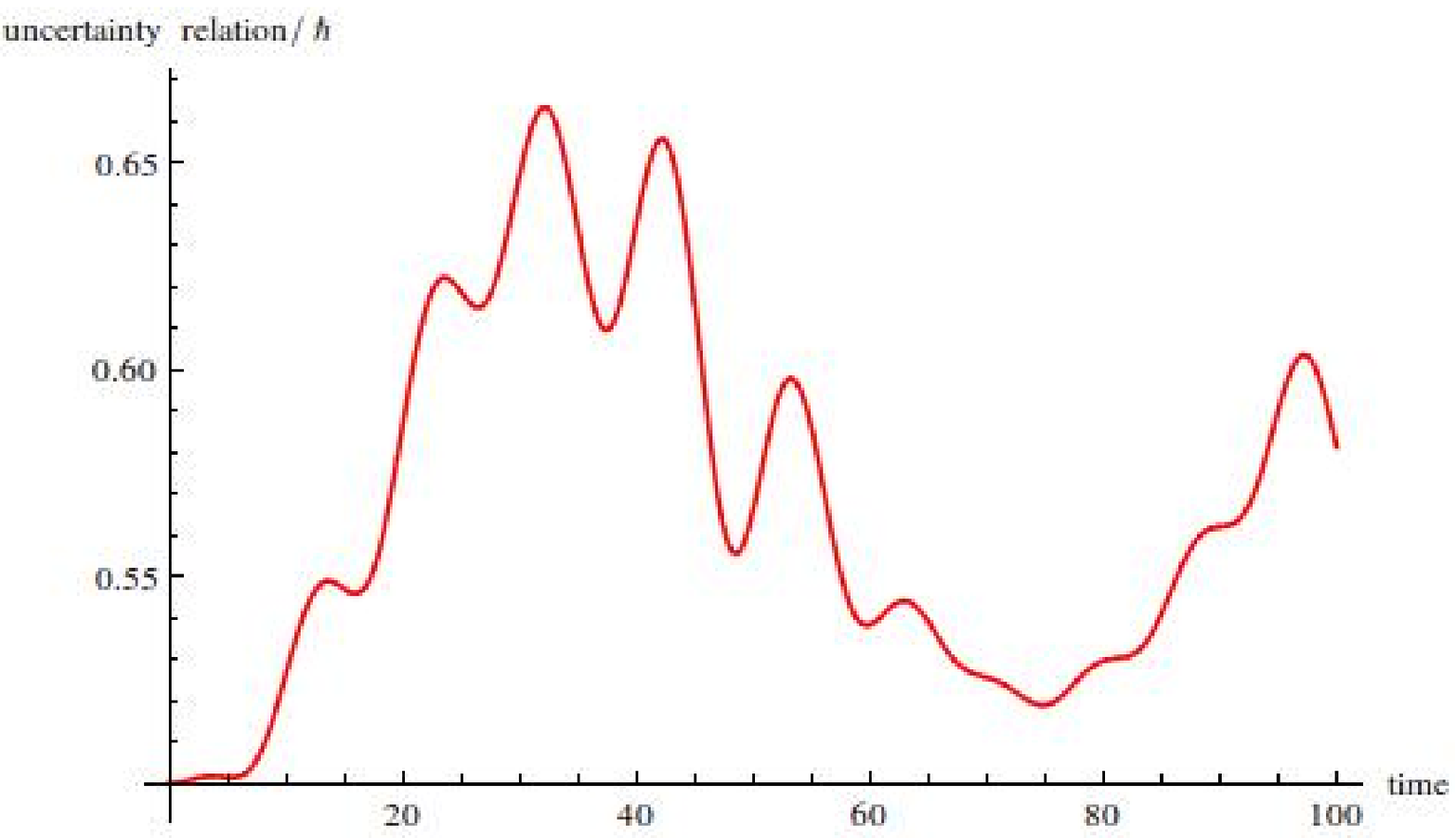}
 \caption{ The time evolution of the uncertainty relation  $\Delta\xi\Delta\rho/\hbar $ for $r=0.1$(blue curve) and $r=0.5$ (red curve) for Linear coherent states of system 1. in the two graphs we take $\varphi=0$ and $\frac{b}{\hbar}=1$.}
 \end{figure}
 \end{center}
Analysing figure 2, we see that the uncertainty is also localized for small $r$. If we compare  figure $2$ with figure $1$, we can conclude that the uncertainty on the linear coherent state is more localized than the uncertainty on the GHA coherent state for a given $r$.
\subsection{Spectrum type 2}
We repeat the same calculations shown in Section (5.1) for the GHA coherent (see [5])\begin{equation}
 \left\vert r,\varphi \right\rangle=N(r)\sum_{n=1}^{\infty}n r^{(n-1)}e^{i(n-1)\varphi}\left\vert n\right\rangle,
\end{equation} of a system whose energy spectrum is 
 \begin{equation}
 \varepsilon_{n}=b\frac{n^{2}}{(n+1)^{2}},
 \end{equation}
for $n\geq0$, where $N(r)=\sqrt{\frac{(1-r^2)^{3}}{1+r^{2}}}$,  with $0\leq r\leq 1$ and $b$ is some constant which has the dimension of energy.\\
\subsubsection*{Time evolutions of the mean values of $\xi$,$\rho$,$\xi^{2}$ and $\rho^{2}$ for the GHA coherent state Eq. $(53)$}
Now, the time evolutions of the mean values of the operators $\xi$ , $\rho$, $\xi^{2}$ and $\rho^{2}$  become
\begin{equation}
<\xi(t)>={L\sqrt{2}}(N(r))^{2}\lbrace \sum_{n=1}^{\infty}r^{2n-1}n(n+1)\sqrt{n}\cos((\frac{n^2}{(n+1)^2}-\frac{(n+1)^2}{(n+2)^2})b\frac{t}{\hbar}+\varphi)\rbrace,
\end{equation}
\begin{equation}
<\rho(t)>=\frac{\sqrt{2}\hbar}{L}(N(r))^{2}\lbrace \sum_{n=1}^{\infty}r^{2n-1}n(n+1)\sqrt{n}\sin((\frac{n^2}{(n+1)^2}-\frac{(n+1)^2}{(n+2)^2})b\frac{t}{\hbar}+\varphi)\rbrace,
\end{equation}
\begin{align}\label{A_Label}
\begin{split}
<\xi(t)^{2}>=&L^{2}(N(r))^{2}\lbrace \sum_{n=1}^{\infty}r^{2n}n(n+2)\sqrt{n(n+1)}\cos((\frac{n^2}{(n+1)^2}-\frac{(n+2)^2}{(n+3)^2})b\frac{t}{\hbar}+2\varphi)\\&\quad+\sum_{n=1}^{\infty}r^{2(n-1)}n^{2}(n-1\rbrace+\frac{L^{2}}{2},\end{split}
\end{align}
\begin{align}\label{A_Label}
\begin{split}
<\rho(t)^{2}>=& -\frac{\hbar^{2}}{L^{2}}(N(r))^{2}\lbrace \sum_{n=1}^{\infty}r^{2n}n(n+2)\sqrt{n(n+1)}\cos((\frac{n^2}{(n+1)^2}-\frac{(n+2)^2}{(n+3)^2})b\frac{t}{\hbar}+2\varphi)\\&\quad-\sum_{n=1}^{\infty}r^{2(n-1)}n^{2}(n-1\rbrace+\frac{\hbar^{2}}{2L^{2}}.\end{split}
\end{align}

\subsubsection*{ Time evolutions of the mean values of $\xi$,$\rho$,$\xi^{2}$ and $\rho^{2}$ for the linear coherent state}
The respective time evolutions of the mean values of the operators $\xi$, $\rho$, $\xi^{2}$ and $\rho^{2}$ for the linear coherent state associated with the system whose energy spectrum is given by Eq. $(54)$ are
 \begin{equation}
<\xi(t)>={L\sqrt{2}}e^{-r^{2}}\lbrace \sum_{n=1}^{\infty}\frac{r^{2n-1}}{(n-1)!}\cos((\frac{n^2}{(n+1)^2}-\frac{(n+1)^{2}}{(n+2)^2})b\frac{t}{\hbar}+\varphi)\rbrace,
\end{equation}
\begin{equation}
<\rho(t)>=\frac{\sqrt{2}\hbar}{L}e^{-r^{2}}\lbrace \sum_{n=1}^{\infty}\frac{r^{2n-1}}{(n-1)!}\sin((\frac{n^2}{(n+1)^2}-\frac{(n+1)^2}{(n+2)^2})b\frac{t}{\hbar}+\varphi)\rbrace,
\end{equation}
\begin{equation}
<\xi(t)^{2}>=L^{2}\lbrace e^{-r^{2}}\sum_{n=1}^{\infty}\frac{r^{2n}}{(n-1)!}\cos((\frac{n^2}{(n+1)^2}-\frac{(n+2)^2}{(n+3)^2})b\frac{t}{\hbar}+2\varphi)+r^{2}\rbrace+\frac{L^{2}}{2},
\end{equation}
\begin{equation}
<\rho(t)^{2}>= -\frac{\hbar^{2}}{L^{2}}\lbrace e^{-r^{2}} \sum_{n=1}^{\infty}\frac{r^{2n}}{(n-1)!}\cos((\frac{n^2}{(n+1)^2}-\frac{(n+2)^2}{(n+3)^2})b\frac{t}{\hbar}+2\varphi)-r^{2}\rbrace+\frac{\hbar^{2}}{2L^{2}}.
\end{equation}
Then, if we project Eqs. $(55)-(58)$ and Eqs. $(59)-(62)$ in Eq. $(34)$ we can calculate the time evolutions of the uncertainty relation for the  GHA coherent state and the linear coherant state respectively.
\begin{center}    
 \begin{figure}[H]
  \includegraphics[scale=0.5]{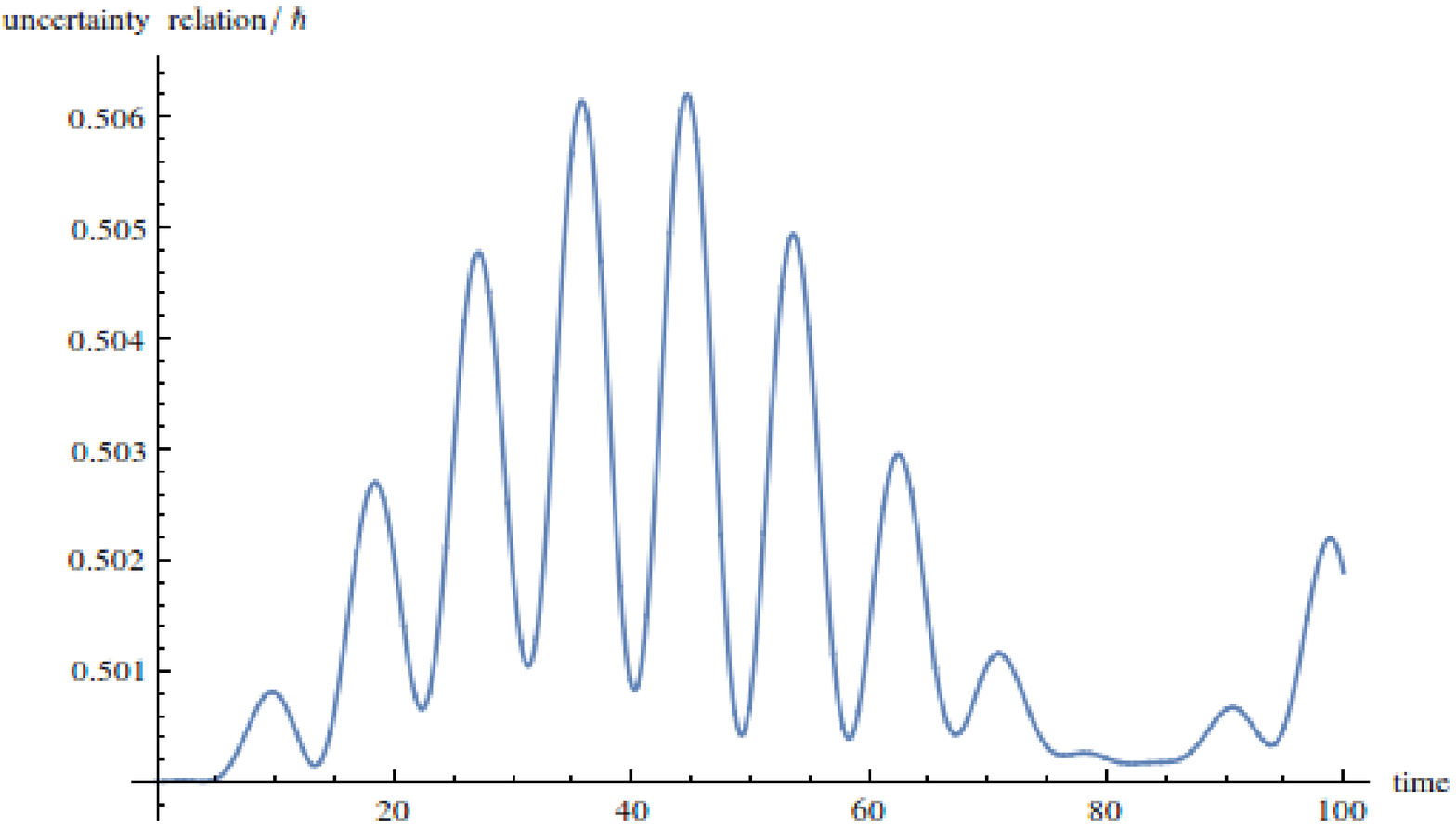}
 \includegraphics[scale=0.5]{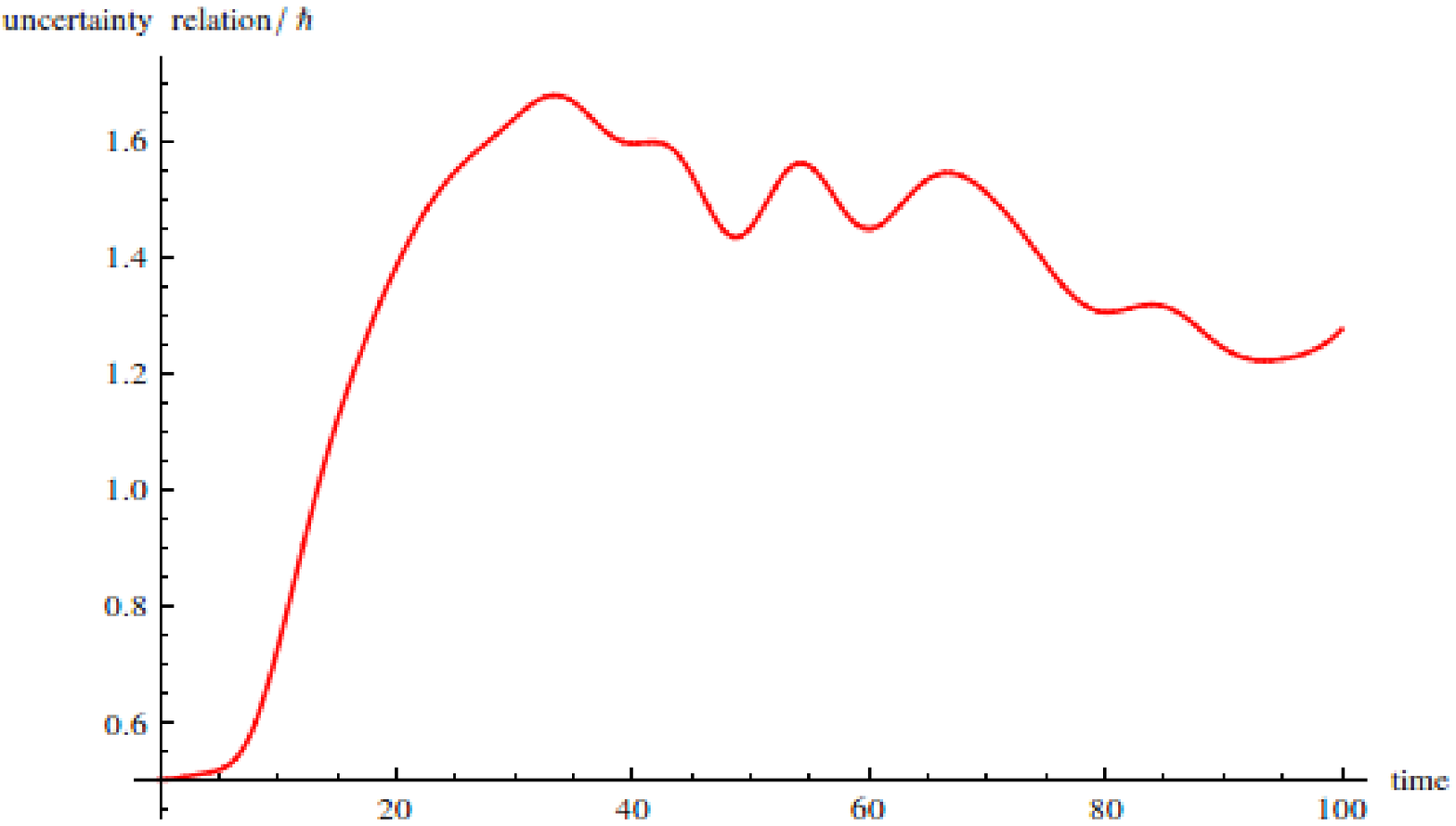}   
  \caption{ The time evolution of the uncertainty relation  $\Delta\xi\Delta\rho/\hbar $ for $r=0.1$(blue curve) and $r=0.5$ (red curve) for GHA coherent states of system 2. In the two graphs we take $\varphi=0$ and $\frac{b}{\hbar}=1$.  }
  \end{figure}
  \end{center}
  \begin{center}    
 \begin{figure}[H]
  \includegraphics[scale=0.5]{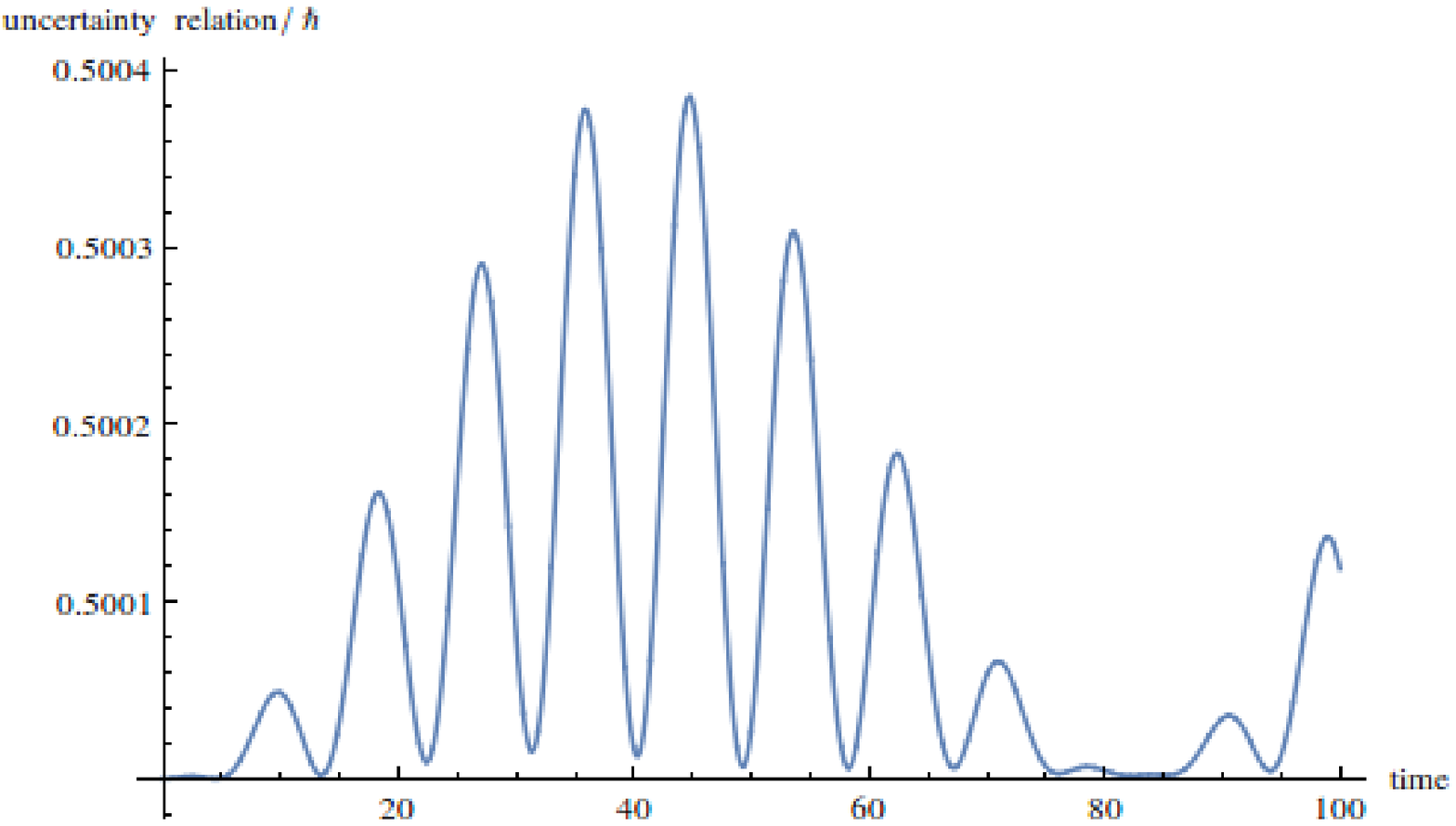}
 \includegraphics[scale=0.5]{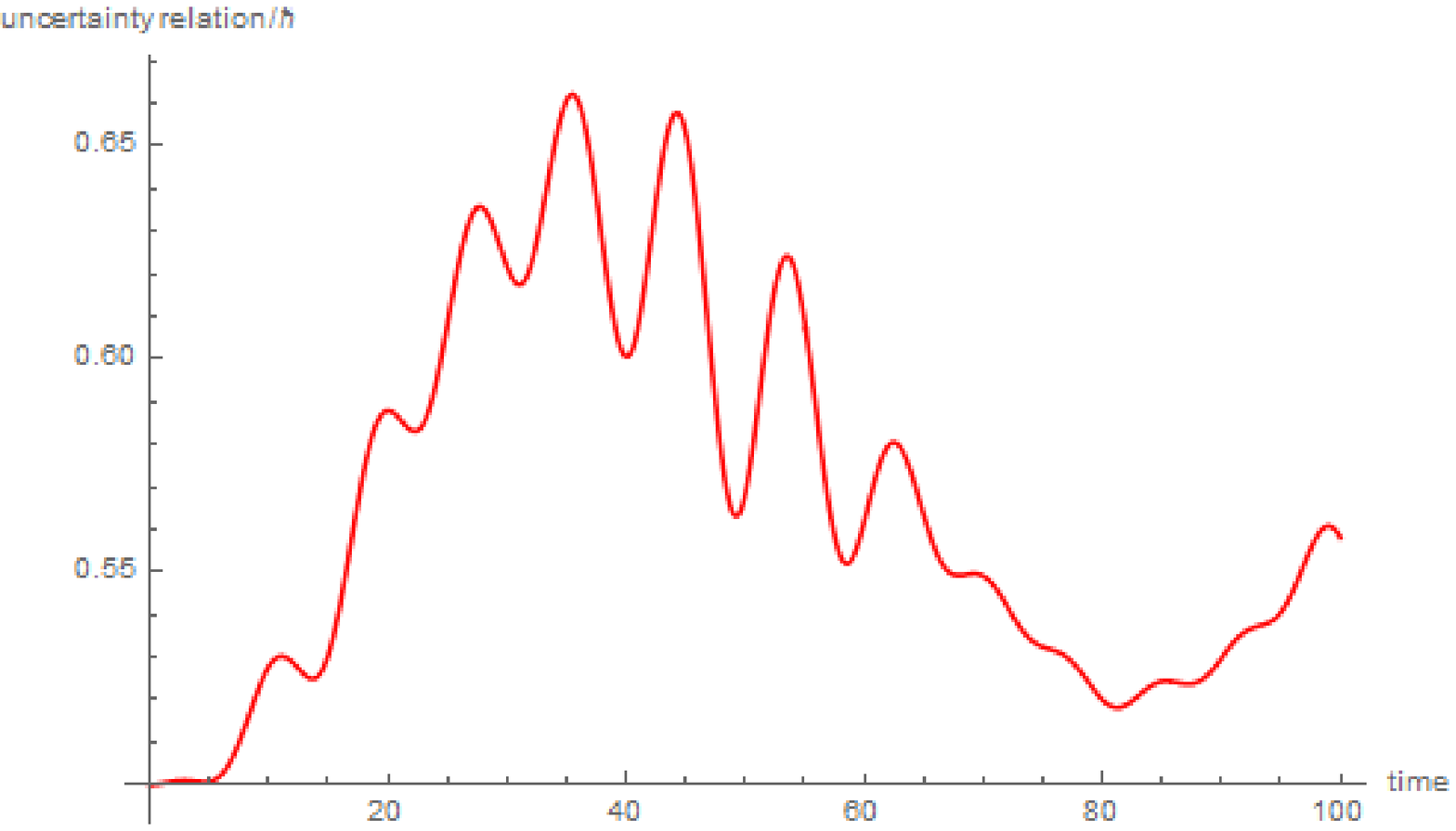} 
  \caption{  The time evolution of the uncertainty relation  $\Delta\xi\Delta\rho/\hbar $ for $r=0.1$(blue curve) and $r=0.5$ (red curve) for Linear coherent states of system 2. In the two graphs we take $\varphi=0$ and $\frac{b}{\hbar}=1$.  }
 \end{figure} \end{center}
 Comparing the different curves of figure 3 and figure 4 , we can conclude that the time uncertainty relation is more localized in the case of liner coherent states compared to the case of GHA coherent states, and we can see that the uncertainty is close to $\frac{\hbar}{2}$ when $r$ is very small for two kinds of coherent states. 
\subsection{Hydrogen atom}
The aim of this section is to calculate the time evolution of the uncertainty relation for the linear and nonlinear coherent states of the Hydrogen atom. In [7] the time evolution of the coherent state of the Hydrogen atom  has been given as 
\begin{equation}
 \left\vert r,\varphi \right\rangle=N(r)\sum_{n=1}^{\infty}\frac{nr^{(n-1)}e^{-i\varepsilon_{n}\frac{t}{\hbar} +i(n-1)\varphi}}{\prod_{i=1}^{n-1}\sqrt{\epsilon_{i+1}-\epsilon_{1}}}\left\vert n \right\rangle,
\end{equation}
where \begin{equation}
N(r)=\sqrt{\frac{r^4(1-r^2)^3}{2r^2(1-2r^2+3r^4)+2(1-r^{2})^{3}\ln(1-r^{2})}},
\end{equation}

is the normalisation constant,  $\varepsilon_{n}=\frac{-b}{n^{2}}$, and $\epsilon_{i}=\frac{\varepsilon_{i}}{b}$ where $b=E_{I}$  is the ionization energy of the Hydrogen atom. We recall that the parameter space, in this case, satisfies $0\leq r\leq 1$.\\ We considered $\left\vert n \right\rangle =(1/n)\sum_{l=0}^{n-1}\sum_{m=-l}^{l}\left\vert n,l,m \right\rangle $ where $n,l$ and $m$ are the quantum numbers which characterize the Hydrogen atom.\\It was verified in [7] that the Hydrogen atom (Eq. ($63$)) behaves as a classical system when the parameter space $r$ is very small.
\subsubsection*{Time evolution of the mean values of $\xi$,$\rho$,$\xi^{2}$ and $\rho^{2}$ for the GHA coherent state of the Hydrogen atom} When we repeat the calculations as shown in previous sections to calculate the mean values of $\xi$, $\rho$, $\xi^{2}$ and $\rho^{2}$ of the coherent state Eq. $(63)$ we find that 
\begin{equation}
<\xi(t)>=\sqrt{2}L (N(r))^{2}\sum_{n=1}^{\infty}\frac{2n^{2}(n+1)r^{2n-1}}{\sqrt{n+2}}\cos((- \frac{1}{n^{2}}+\frac{1}{(n+1)^{2}})b\frac{t}{\hbar}+\varphi),
\end{equation}
 \begin{equation}
<\rho(t)>=\frac{\sqrt{2}\hbar}{L} (N(r))^{2}\sum_{n=1}^{\infty}\frac{2n^{2}(n+1)r^{2n-1}}{\sqrt{n+2}}\sin((- \frac{1}{n^{2}}+\frac{1}{(n+1)^{2}})b\frac{t}{\hbar}+\varphi),
\end{equation}
\begin{align}\label{A_Label}
\begin{split}
<\xi(t)^{2}>=&L^{2} (N(r))^{2}\lbrace\sum_{n=1}^{\infty}\frac{2n^{2}(n+2)\sqrt{n+2}r^{2n}}{\sqrt{n+3}}\cos((- \frac{1}{n^{2}}+\frac{1}{(n+2)^{2}})b\frac{t}{\hbar})+2\varphi)\\&\quad+\sum_{n=1}^{\infty}\frac{2n^{3}(n-1)r^{(2n-2)}}{n+1}\rbrace+\frac{L^{2}}{2},\end{split}
\end{align}
and 
\begin{align}\label{A_Label}
\begin{split}
<\rho(t)^{2}>=&-\frac{\hbar^{2}}{L^{2}}(N(r))^{2}\lbrace\sum_{n=1}^{\infty}\frac{2n^{2}(n+2)\sqrt{n+2}r^{2n}}{\sqrt{n+3}}\cos((- \frac{1}{n^{2}}+\frac{1}{(n+2)^{2}})b\frac{t}{\hbar})+2\varphi)\\&\quad-\sum_{n=1}^{\infty}\frac{2n^{3}(n-1)r^{(2n-2)}}{n+1}\rbrace+\frac{\hbar^{2}}{2L^{2}}.\end{split}
\end{align}
\subsubsection*{ Time evolution of the mean values of $\xi$,$\rho$,$\xi^{2}$ and $\rho^{2}$ for the linear coherent states of the Hydrogen atom}
Now, we calculate the mean values of the operators $\xi$, $\rho$, $\xi^{2}$ and $\rho^{2}$ for the linear coherent state corresponding to the Hydrogen atom
 \begin{equation}
<\xi(t)>={L\sqrt{2}}e^{-r^{2}}\lbrace \sum_{n=1}^{\infty}\frac{r^{2n-1}}{(n-1)!}\cos((- \frac{1}{n^{2}}+\frac{1}{(n+1)^{2}})b\frac{t}{\hbar}+\varphi)\rbrace,
\end{equation}
\begin{equation}
<\rho(t)>=\frac{\sqrt{2}\hbar}{L}e^{-r^{2}}\lbrace \sum_{n=1}^{\infty}\frac{r^{2n-1}}{(n-1)!}\sin((- \frac{1}{n^{2}}+\frac{1}{(n+1)^{2}})b\frac{t}{\hbar}+\varphi)\rbrace,
\end{equation}
\begin{equation}
<\xi(t)^{2}>=L^{2}\lbrace e^{-r^{2}} \sum_{n=1}^{\infty}\frac{r^{2n}}{(n-1)!}\cos((- \frac{1}{n^{2}}+\frac{1}{(n+2)^{2}})b\frac{t}{\hbar}+2\varphi)+r^{2}\rbrace+\frac{L^{2}}{2},
\end{equation}
\begin{equation}
<\rho(t)^{2}>= -\frac{\hbar^{2}}{L^{2}}\lbrace e^{-r^{2}} \sum_{n=1}^{\infty}\frac{r^{2n}}{(n-1)!}\cos((- \frac{1}{n^{2}}+\frac{1}{(n+2)^{2}})b\frac{t}{\hbar}+2\varphi)-r^{2}\rbrace+\frac{\hbar^{2}}{2L^{2}}.
\end{equation}
\begin{center}    
 \begin{figure}[H] 
 \includegraphics[scale=0.5]{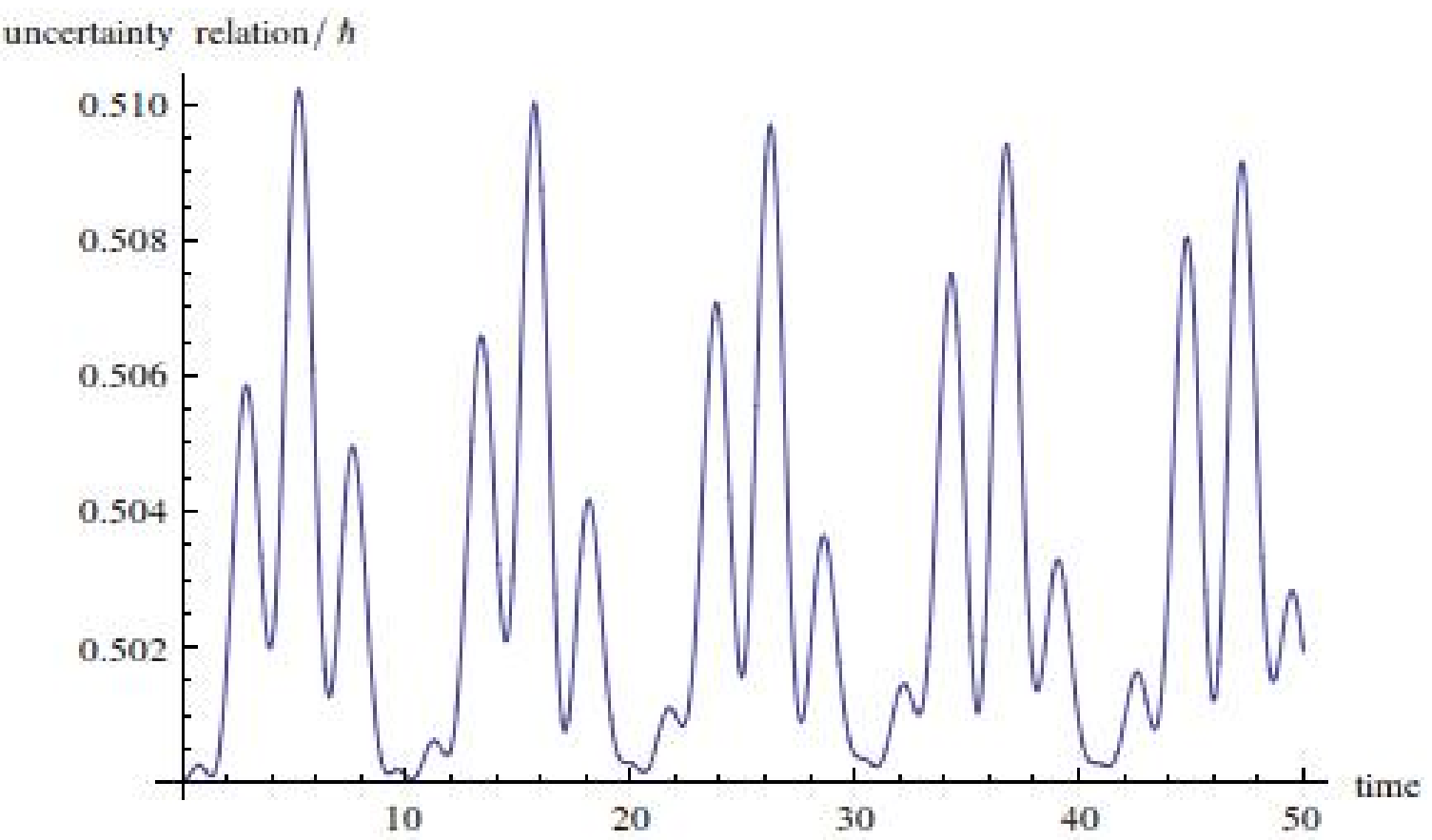}  
  \includegraphics[scale=0.5]{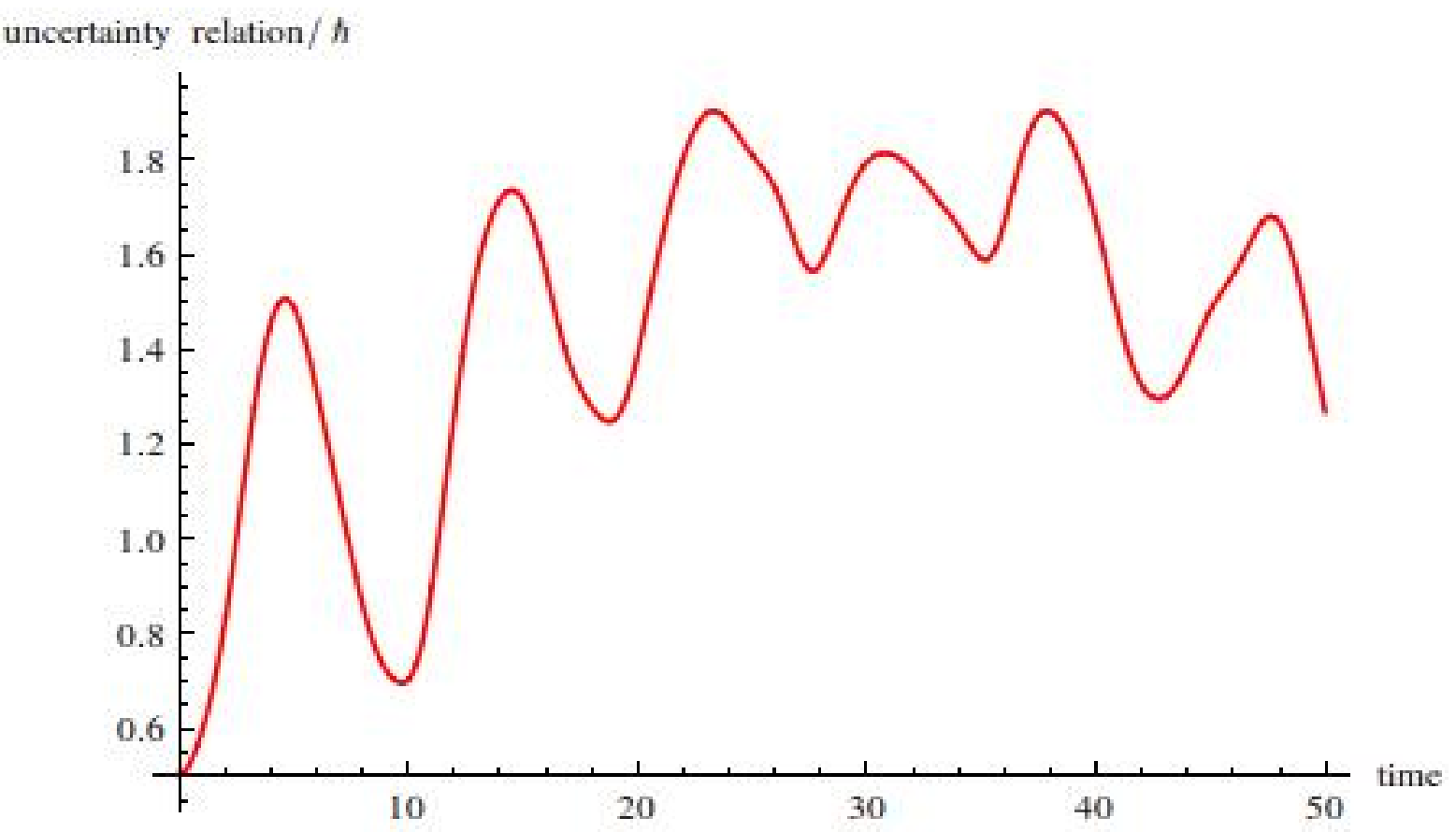}  
  \caption{  The time evolution of the uncertainty relation  $\Delta\xi\Delta\rho/\hbar $ for $r=0.1$(blue curve) and $r=0.5$ (red curve) for GHA coherent states of Hydrogen atom. In the two graphs we take $\varphi=0$ and $\frac{b}{\hbar}=1$.  }
  \end{figure}
  \end{center}
\begin{center}    
 \begin{figure}[H]
  \includegraphics[scale=0.5]{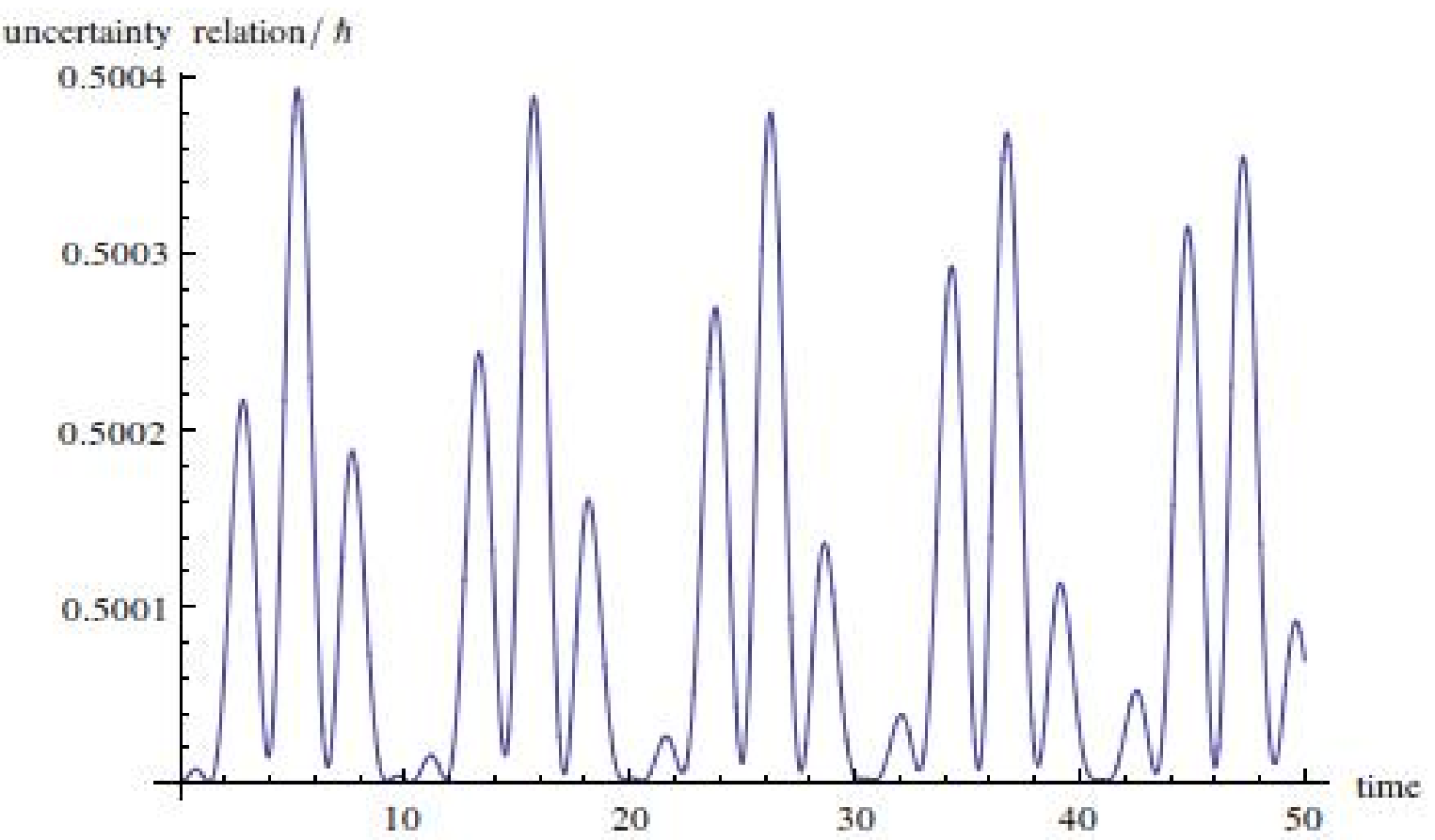}
 \includegraphics[scale=0.5]{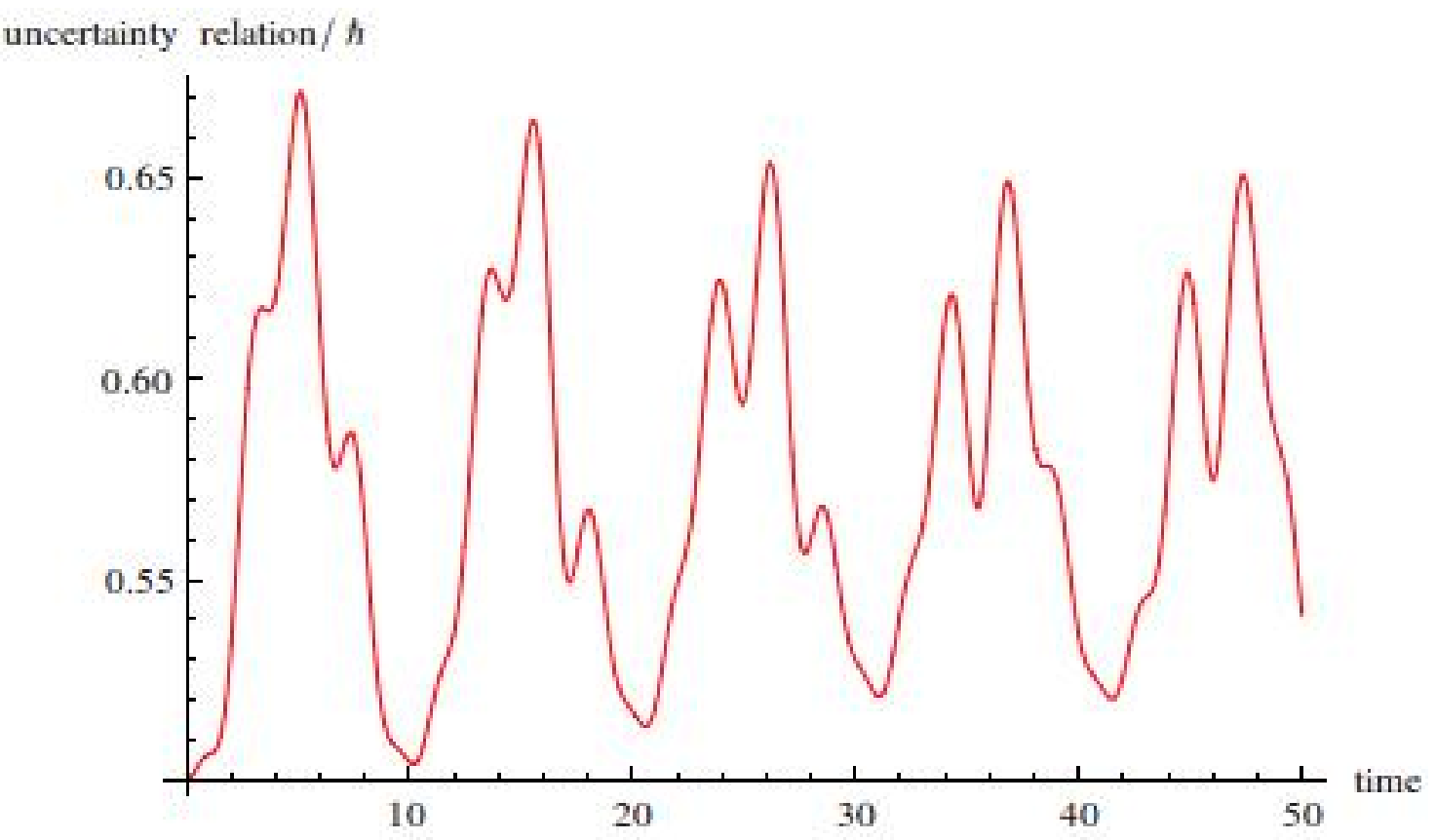}   
  \caption{ The time evolution of the uncertainty relation  $\Delta\xi\Delta\rho/\hbar $ for $r=0.1$(blue curve) and $r=0.5$ (red curve) for linear coherent states of Hydrogen atom. In the two graphs we take $\varphi=0$ and $\frac{b}{\hbar}=1$.  }
  \end{figure}
  \end{center}
The time evolutions of the uncertainty relation of the GHA coherent state and the linear coherent state of the Hydrogen atom can be reached by using Eqs. (65)-(68) and Eqs. (69)-(72) respectively. They are plotted in figure 5 and figure 6 for two different parameter spaces $r=0,1$ and $r=0,5$. Looking at these two figures we can see easily that the time evolutions are more localized when $r=0,1$ for two kinds of coherent states. It follows then that when the space parameter is very small the system behaves as a classical system, this property is in accordance with the results found in [5]. Another important note is that the time evolution of the uncertainty  is more localized for the linear coherent state than  the time evolution of the uncertainty of the GHA coherent state.
 \section{ Morse potential from GHA}
\subsection{Nilpotency}
In this section, we give a brief definition of nilpotency. Let us concider a vector space $\mathcal{H}$ and an operator $A$  which acts on $\mathcal{H}$  . $A$ is a nilpotent operator if there exists an integer number $m\geq1$, such that $A^{m}=0$.\\We define the index of nilpotence s by the formula \begin{equation}
s=Min\lbrace m\geq 1 \hspace*{0.1cm}|\hspace*{0.1cm}A^{m}=0\rbrace.
\end{equation}
This leads to  $A^{s}=0$, with $A^{s-1}\ne 0$.\\
The subspace of $\mathcal{H}$ generated by $\lbrace$$\left\vert n \right\rangle$, $A\left\vert n \right\rangle$, $A^{2}\left\vert n \right\rangle$,..., $A^{s-1}\left\vert n \right\rangle$$\rbrace$ is called the cyclic subspace associated with $A$ , where $\left\vert n \right\rangle$ is a vector that belongs  to $\mathcal{H}$. In this way, one can show that \\$( \left\vert n \right\rangle$, $A\left\vert n \right\rangle$, $A^{2}\left\vert n \right\rangle$,..., $A^{s-1}\left\vert n \right\rangle)$ is a basis of the cyclic subspace[16].\\
In the following section, we will show that GHA of the Morse potential is a nilpotent algebra by giving the corresponding characteristic function. Moreover, we will show that its representation dimension must be finite .
 \subsection{Morse potential system}
We recall that the Morse potential is the best approximation to describe the vibrations inside a diatomic molecule and it also appears in the
spectroscopy of diatomic molecules and anharmonic vibrational dynamics [17]. The one-dimensional Morse model is given by Shr$\ddot{o}$dinger equation
\begin{equation}
H\psi(x)=\left(\frac{\hat{P}^{2}}{2m_{r}}+V_{0}(e^{-2\beta x}-2e^{-\beta x})\right)\psi(x)=E\psi(x),
\end{equation}
where  $ x $  is the displacement of the two atoms from their equilibrium positions. $V_{0}$
 is the depth of the potential well at the equilibrium $x=0$, $\beta$ is the width of the potential and $m_{r}$  is the reduced mass of the oscillating system composed by two atoms of masses $m_{1}$ and  $m_{2}$.\\
The well known energy spectrum is given by 
\begin{equation}
E_{n}=-\frac{\hbar^{2}\beta^{2}}{2m_{r}}(p-n)^{2},
\end{equation}
where 
\begin{equation}
p=\frac{\nu-1}{2},\hspace{1cm}\nu=\sqrt{\frac{8m_{r}V_{0}}{\beta^{2}\hbar^{2}}},
\end{equation}
We notice that the spectrum is finite  $\lbrace n=0, 1, 2, ..., [p]\rbrace$   with $[p]$ is the integer part of $p$.\\
The energy eigenfunctions are given by 
\begin{equation}
\psi_{n}^{\nu}(y)=\mathcal{N}_{n}e^{-\frac{y}{2}}y^{s}L_{n}^{2s}(y),
\end{equation}
 where we have used the change of variable  $y=\nu e^{-\beta x}$, and $L_{n}^{2s}(y)$ are the associated Laguerre functions with $2s=\nu-2n-1$ and
$ \mathcal{N}_{n} $ is the normalization constant given by 
\begin{equation}
\mathcal{N}_{n}=\sqrt{\frac{\beta(\nu-2n-1)\Gamma(n+1)}{\Gamma(\nu-n)}}.
\end{equation} 
 Now we are in a position to install the action of the GHA generators.
\subsection{Action of $ A $, $ A^{\dagger} $and  $ J_{0} $ }
As mentioned in section 2 and in [5], the generator $J_{0}$  is a hermitian operator and we can take $J_{0}$ as the Hamiltonian of the physical system under consideration. In the following we will take $ J_{0}=H/(\frac{\hbar^{2}\beta^{2}}{2m_{r}}) $.\\ consequently, the eigenvalues of $J_{0}$ are $ \varepsilon_{n}=E_{n}/(\frac{\hbar^{2}\beta^{2}}{2m_{r}}) $.
Let us act by $ A $  and $ A^{\dagger} $ on a vector $ \left\vert n \right\rangle $ of the Fock space. Owing to the Eq. (75), one can prove that for $n=0,1,2,...[p]-1$
\begin{equation}
\varepsilon_{n+1}-\varepsilon_{0}=(n+1)(2p-n-1).
\end{equation}
It follows from Eq. $(15)$ that one can show the following expression for  $n=0, 1, 2,..., [p]$ \begin{equation}
 N_{n}^{2}=(n+1)(2p-n-1).
\end{equation}
Consequently, we find that for  $n=0, 1, 2,..., [p]$ the expression 
\begin{equation}
A\left\vert n \right\rangle=\sqrt{n(2p-n)}\left\vert n-1 \right\rangle.
\end{equation}
However, the action 
\begin{equation}
A^{\dagger}\left\vert n \right\rangle=\sqrt{(n+1)(2p-n-1)}\left\vert n+1 \right\rangle.
\end{equation}   
is valid only for $n=0, 1, 2,..., [p]-1$. \\

From Eq. $(81)$  we have the vacuum state condition $ A\left\vert 0 \right\rangle=0$ , and as $A$ and $A^{\dagger}$ are each other's adjoint, we can easily conclude that 
\begin{equation}
 A^{\dagger}\left\vert n_{max} \right\rangle=0 \hspace{0.5 cm}  where \hspace{0.5 cm} n_{max}=[p].
 \end{equation} 
 Consequently, $(A^{\dagger})^{n_{max}+1}=0$. Thus, $A^{\dagger}$ is a
nilpotent operator (see section 6.2) otherwise we lose the vacuum
state condition, showing that GHA, in this case, is a nilpotent algebra, and the representation is finite. $n_{max}+1$ is the index of the nilpotency. In this case, the Fock
space may be seen as a periodic space.  \\
The operator $J_{0}$ in terms of $A$ and $A^{\dagger}$ is given by 
\begin{equation}
J_{0}=A^{\dagger}A-p^{2}.
\end{equation}
\subsection{The characterstic function of the algebra}
In this section, we present one of the main result of the paper. we show the appropriate characteristic function for the Morse oscillator and we give the commutation relations which connect between the GHA generators.  
From Eq. $(75)$, the energy levels are given by 
\begin{equation}
\varepsilon_{n+1}=-(p-(n+1))^{2},
\end{equation}
for  $n<n_{max}$. We can easily see in this case that
\begin{equation}
\varepsilon_{n+1}=\varepsilon_{n}+2\sqrt{-\varepsilon_{n}}-1.
\end{equation}
Therefore, we can conclude that the characteristic function can be written for any  $n<n_{max}$ as follows 
\begin{equation}
f(x)=x+2\sqrt{-x}-1.
\end{equation}
However, to get $A^{\dagger}\left\vert n_{max} \right\rangle=0$  we should take $f(\varepsilon_{n_{max}})=\varepsilon_{0}$ .\\Accordingly, the nilpotency has added a restriction on the characteristic function. 
 The GHA generated by $J_{0}$ , $A$ and $A^{\dagger}$ becomes from  Eqs. $(1)-(2)-(3)-(87)$:

\begin{equation}
   [J_{0},A^{\dagger}]=2A^{\dagger}\sqrt{-J_{0}}-A^{\dagger},
\end{equation}
\begin{equation}
[J_{0},A]=-2\sqrt{-J_{0}}A+A,
\end{equation}
\begin{equation}
   [A,A^{\dagger}]=2\sqrt{-J_{0}}-I ,
\end{equation}
where $I $ is the identity operator.
\subsection{Morse coherent state }
In this section, we construct the coherent states corresponding to the Morse oscillator, Applying the relation  Eq. $(27)$ to our case, since the spectrum of the Morse potential is finite, we find the corresponding coherent state
 \begin{equation}
 \left\vert z \right\rangle=N(|z|)\sum_{n=0}^{[p]-1}\frac{z^{n}}{\sqrt{n!\prod_{i=1}^{n}(2p-i)}}\left\vert n \right\rangle.
 \end{equation}
The normalization function is given by
 \begin{equation}
 N(|z|)=(\sum_{n=0}^{[p]-1}\frac{|z|^{2n}}{n!\prod_{i=1}^{n}(2p-i)})^{-1/2}.
 \end{equation}
 
The time evolution of the coherent state (Eq. $(91)$) is obtained by the application of the unitary operator Eq. $(33)$

\begin{equation}
\left\vert z(t) \right\rangle=U\left\vert z \right\rangle=N(z)\sum_{n=0}^{[p]-1}\frac{z^{n}e^{\frac{-iHt}{\hbar}}}{\sqrt{n!\prod_{i=1}^{n}(2p-i)}}\left\vert n \right\rangle ,
\end{equation}
\begin{equation}
 \hspace{2.1cm} =N(z)\sum_{n=0}^{[p]-1}\frac{z^{n}e^{\frac{i\hbar\beta^{2}(p-n)^2t}{2m_{r}}}}{\sqrt{n!\prod_{i=1}^{n}(2p-i)}}\left\vert n \right\rangle.
\end{equation}
\subsection{Time evolution of the uncertainty relation $\Delta\xi\Delta\rho$ for GHA coherent state of Morse oscillator}
This section aims to calculate the time evolution of the operators $\xi$, $\rho$, $\xi^{2}$ and $\rho^{2}$  on GHA coherent state of the Morse oscillator Eq. $(94)$, Then we can conclude the time evolution of the uncertainty relation Eq. $(34)$. First let us recall that the representation, in this case, is finite and the algebra is nilpotent $(A^{\dagger})^{n_{max}+1}=0$, Consequently the corresponding generalized harmonic oscillator creation and annihilation operators $D$ and $D^{\dagger}$ act on a vector $\left\vert n \right\rangle$ with $n=0,1,...,n_{max}-1$, as Eqs. $(18-19)$. However in this case 
\begin{equation}
D^{\dagger}\left\vert n_{max} \right\rangle=0.
\end{equation}
Applying Eqs. $(18)-(19)-(23)-(24)-(92)-(94)$ we find that 
\begin{equation}
<\xi(t)>=\sqrt{2}L (N(r))^{2}\sum_{n=1}^{n_{max}-1}\frac{r^{2n+1}\sqrt{n+1}}{\sqrt{n!(n+1)!\prod_{i=1}^{n}(2p-i)\prod_{i=1}^{n+1}(2p-i)}}\cos((2(n-p)+1)\frac{\hbar\beta^{2}}{2m_{r}}t+\varphi),
\end{equation}
 \begin{equation}
<\rho(t)>=\frac{\sqrt{2}\hbar}{L} (N(r))^{2}\sum_{n=1}^{n_{max}-1}\frac{r^{2n+1}\sqrt{n+1}}{\sqrt{n!(n+1)!\prod_{i=1}^{n}(2p-i)\prod_{i=1}^{n+1}(2p-i)}}\sin((2(n-p)+1)\frac{\hbar\beta^{2}}{2m_{r}}t+\varphi),
\end{equation}
\begin{align}\label{A_Label}
\begin{split}
<\xi(t)^{2}>=&L^{2} (N(r))^{2}\lbrace\sum_{n=1}^{n_{max}-2}\frac{r^{2n+2}\sqrt{(n+1)(n+2)}}{\sqrt{n!(n+2)!\prod_{i=1}^{n}(2p-i)\prod_{i=1}^{n+2}(2p-i)}}\cos(4(n-p+1)\frac{\hbar\beta^{2}}{2m_{r}}t+2\varphi)\\&\quad+\sum_{n=1}^{n_{max}-1}\frac{nr^{2n}}{n!\prod_{i=1}^{n}(2p-i)}\rbrace+\frac{L^{2}}{2},\end{split}
\end{align}
and 
\begin{align}\label{A_Label}
\begin{split}
<\rho(t)^{2}>=&-\frac{\hbar^{2}}{L^{2}}(N(r))^{2}\{\sum_{n=1}^{n_{max}-2}\frac{r^{2n+2}\sqrt{(n+1)(n+2)}}{\sqrt{n!(n+2)!\prod_{i=1}^{n}(2p-i)\prod_{i=1}^{n+2}(2p-i)}}\cos(4(n-p+1)\frac{\hbar\beta^{2}}{2m_{r}}t+2\varphi)\\&\quad-\sum_{n=1}^{n_{max}-1}\frac{nr^{2n}}{n!\prod_{i=1}^{n}(2p-i)}\}+\frac{\hbar^{2}}{2L^{2}}.\end{split}
\end{align}
In Eqs. (98)-(99)  we have considered that the last terms of the first summations do not contribute in calculations, Consequently these summations run from 0 to $n_{max}-2$ because we have, in this case, the important property $D^{\dagger}\left\vert n_{max} \right\rangle=0$. 
\subsubsection*{Application: coherent states for $O_{2}$}
Now let us apply Eqs. $(96-99)$ to simulate the time evolution of the uncertainty relation for the GHA coherent state of the molecule $O_{2}$ which is described by the Morse parametrs  $\beta=2.78\times10^{10}\hspace{0.2cm}m^{-1}\hspace{0.2cm}$   and $V_{0}=5.211\hspace{0.2cm}ev$ (see [18]).The reduced mass of the molecule $O_{2}$ is $m_{r}=1.33 \times10^{-26} \hspace{0.2cm}kg$.  Consequently, $\nu=16.18$ and $n_{max}=7$.
\begin{center}    
 \begin{figure}[H]
  \includegraphics[scale=0.5]{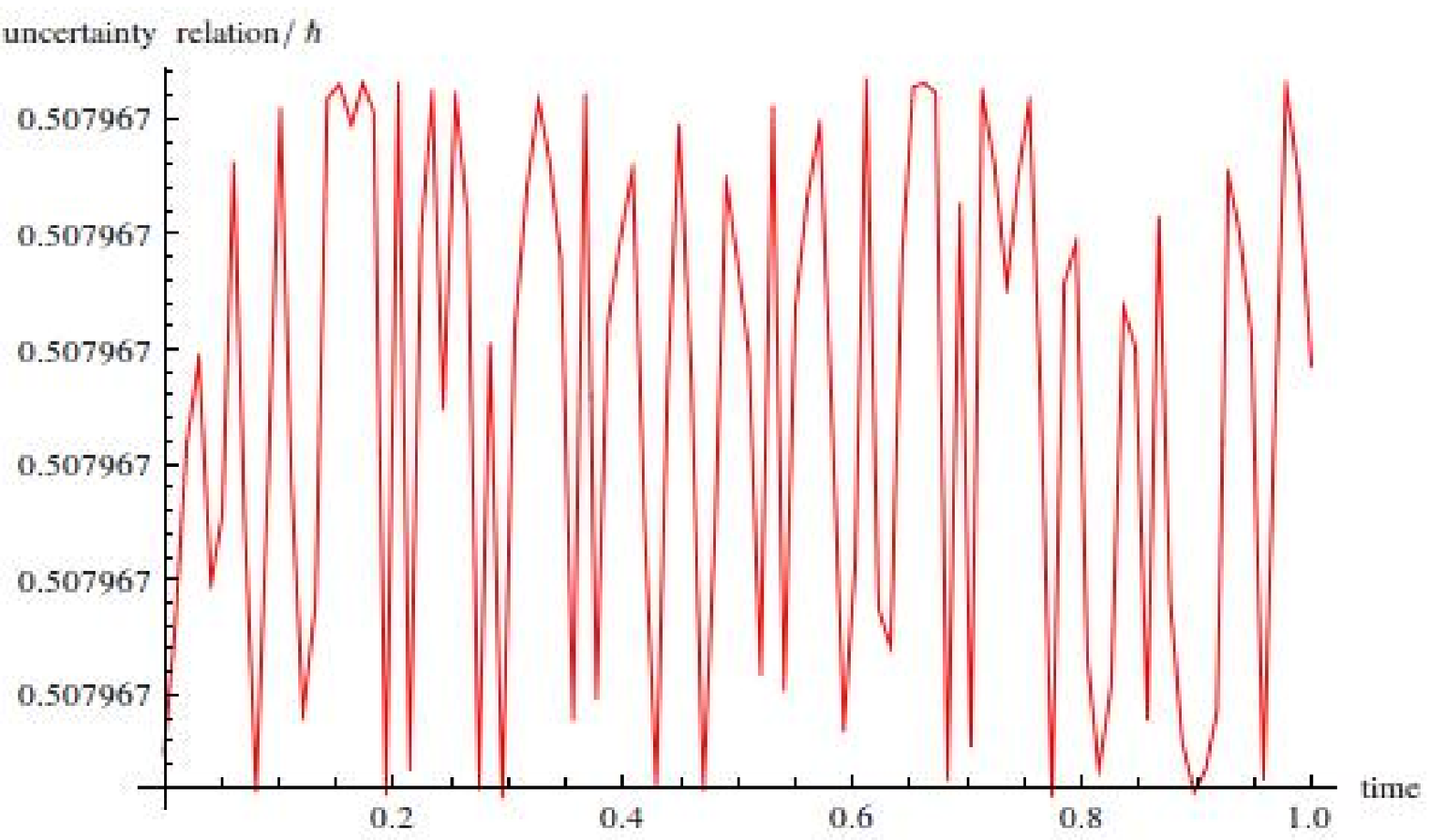}
 \includegraphics[scale=0.5]{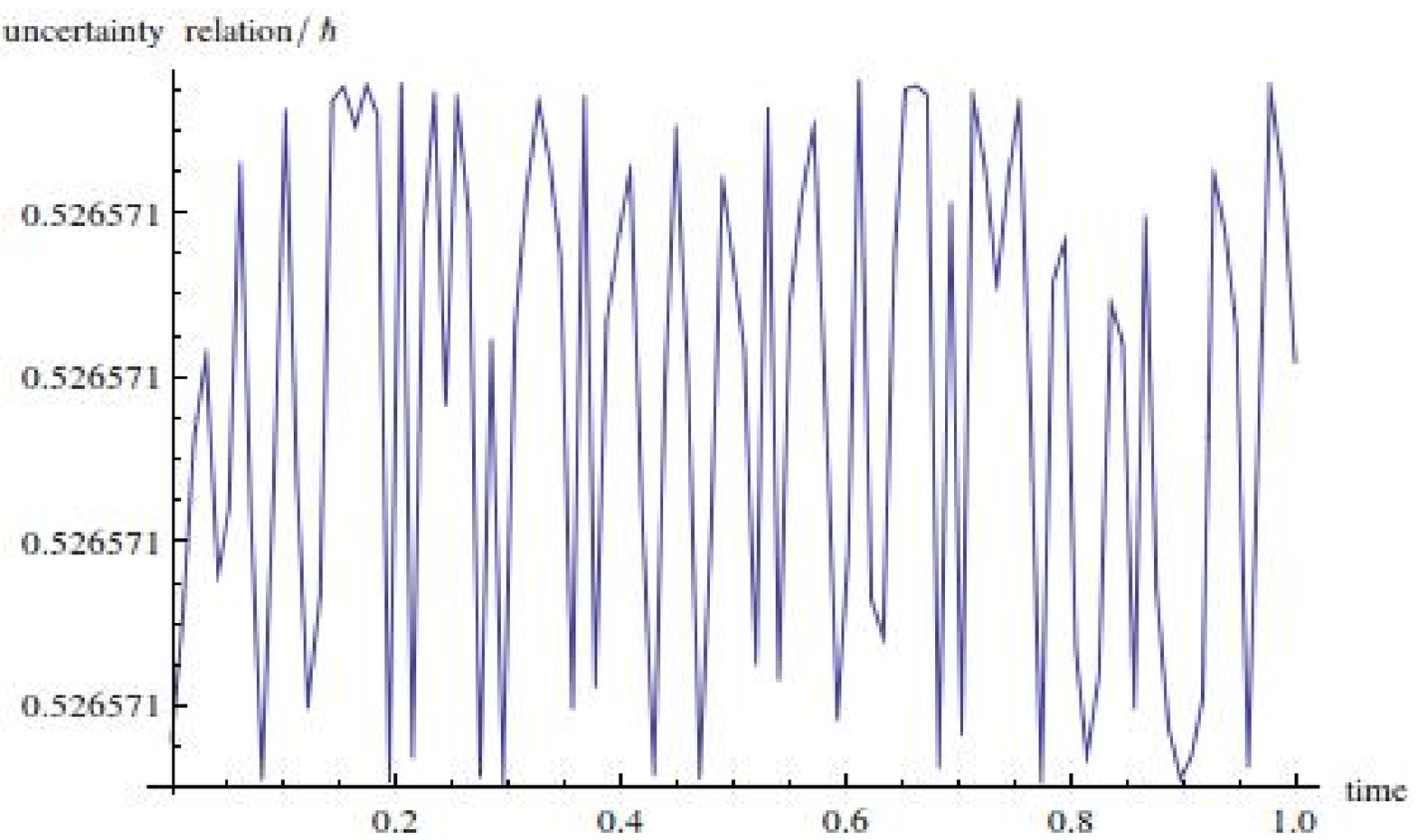}
  \caption{ The time evolution of the uncertainty relation  $\Delta\xi\Delta\rho/\hbar $ for $r=0.03$(red curve) and $r=0.1$ (blue curve) for GHA coherent states of the molecule $O_{2}$.  }
  \end{figure}
  \end{center}
Upon studying figure 8 we can see that the localization of the time evolution of the uncertainty is achieved when the space parameter $r$ is very small ($r=0.03$). We notice that following the approach mentioned in section 2 we can not define the linear coherent state of the Morse oscillator as its spectrum is finite.
\newpage
\section{Conclusion}

In this research paper, we have analyzed the time evolution of the uncertainty relation $\Delta\xi\Delta\rho$ (Eq. $(34)$). We have done this analysis in terms of the operators $D$ and $D^{\dagger}$ which act on a Fock space vector as the harmonic oscillator ladder operators, and on two kinds of coherent states$($the GHA coherent states and the linear coherent states$)$ for particular systems( system 1 and sytem 2) and for the Hydrogen atom. The time evolutions of the uncertainty relation on two kinds of coherent states for different systems $($system 1, system 2 and Hydrogen atom$)$ are close to $\frac{\hbar}{2}$ if the space parameter $r$ is  small. Consequently, the systems behave as the classical systems when the parameter $r$ is very small. For three different systems the time evolution of the uncertainty relation is more localised for the linear coherent states compared to GHA coherent states.\\
In the second part of the paper, we have considered the generalized Heisenberg algebra for the one dimensional Morse oscillator. We have given the representations of this algebra and have shown that the creation generator must be a nilpotent operator for the algebra to be compatible. Subsequently, we have investigated the coherent state for the Morse oscillator, and we have simulated the time evolution of the uncertainy relaion on the coherent state of a system described by the Morse potential (the molecule $O_{2}$). This behaves also as a classical syetem when the space parameter is very small. The quantumness of the Morse potential state will be considered in a future extended work [19] .
\section*{Acknowledgements}
One of authors, Y.H would like to thank the theoretical physics Institut, University of Tubingen, Germany. He would like to thank also the AFFP for finnancial support and for the award.The authors thank A.Belhaj for very interesting comments and for helpful discussions.This work is partially supported by the ICTP through AF-13.
 \newpage
\section*{references}
$\hspace{0.01 cm}$\\
$[1]$ $\hspace{0.1cm}$ M.Arik,D.D.Coon,J.Math.Phys.17(1976)524,L.Biedenharn,J.PhysA 22(1989) l873,A.J.Macfarlane, \\J.Phys.A 22(1989) 4581\\
$[2]$ $\hspace{0.1cm}$D.Bonatsos   C.Daskaloyannis   Prog.Part.Nucl.Phys.43(1999)573.\\
$[3]$ $\hspace{0.1cm}$ M.R. Monteiro, L.M.C.S. Rodrigues, S. Wulk, Phys. Rev. Lett. 76 (1996) 1098.\\
$[4]$ $\hspace{0.1cm}$E.M.F Curado and M. A. rego-Monteiro,J.Phys.A 34,3253(2001)\\
$[5]$ $\hspace{0.1cm}$ Y. Hassouni,E. M. F. Curado and M. A. Rego-Monteiro,Phys. Rev. A 71, 022104 (2005).\\
$[6]$ $\hspace{0.1cm}$E.M.F Curado,Y.Hassouni,M. A. rego-Monteiro,Ligia M.C.S.Rodrigues Physics Letters A 372 (2008) 3350-3355.\\
$[7]$ $\hspace{0.1cm}$E.M.F. Curadoa,A. Rego-Monteiroa, Ligia M.C.S. Rodriguesa, Y. Hassouni Physica A 371 (2006)\\
$[8]$ $\hspace{0.1cm}$ S. Robles-Perez, Y. Hassouni, P.F. Gonzalez-Diaz, Phys.
Lett. B 683, 1 (2010).\\
$[9]$ $\hspace{0.1cm}$E. Shr$\ddot{o}$dinger, Naturwissenschaften 14, 664 (1926).\\
$[10]$ $\hspace{0.1cm}$U. M. and Glauber, R. J., Phys. Rev. 145, 1041 (1966).\\
$[11]$ $\hspace{0.1cm}$A.M. Perelomov, Commun. Math. Phys. 26, 222 (1972).\\
$ [12]$ $\hspace{0.1cm}$A.M. Perelomov, Generalized Coherent States and their Applications (Springer, New York, 1986).\\
$[13]$ J. R. Klauder and B. Skagerstam, Coherent States: Applications in Physics and Mathematical Physics sWorld Scientific, Singapore, 1985d; W.-M. Zhang, D. H. Feng, and R. Gilmore, Rev. Mod. Phys. 62, 867 (1990).\\
$[14]$ $\hspace{0.1cm}$ J. R. Klauder and B. S. Skagertan, Coherent States sWorld Scientific, Singapore, 1985.\\
$[15]$ $\hspace{0.1cm}$M. A. Rego-Monteiro, E. M. F. Curado, and Ligia M. C. S. Rodrigues  Phys. Rev.96, 052122 (2017).\\
$[16]$ $\hspace{0.1cm}$E. Devinatz, M. J. Hopkins, and J. H. Smith. Nilpotence and stable homotopy theory.Annals of Mathematics, 128:207-242, 1988.\\
$[17]$ $\hspace{0.1cm}$P. M. Morse, Phys. Rev. 34, 57-64 (1929).\\
$[18]$ $\hspace{0.1cm}$Daniel D. Konowalow and Joseph O. Hirschfelder  Phys. Fluids 4, 637 (1961).\\$[18]$ $\hspace{0.1cm}$Daniel D. Konowalow and Joseph O. Hirschfelder  Phys. Fluids 4, 637 (1961).\\
$[19]$ $\hspace{0.1cm}$ work is in preparation.\\

\end{document}